\documentclass[11pt,a4paper]{article}

\usepackage{ascmac}
\usepackage{amsmath}
\usepackage{amssymb}
\usepackage{bm}
\usepackage[footnotesize,bf]{caption}
\usepackage{fullpage}
\usepackage{color}
\usepackage{float}
\usepackage{graphicx}
\usepackage[]{natbib}
\usepackage{subfigure}
\usepackage{txfonts}
\usepackage{threeparttable}
\usepackage{wrapfig}
\usepackage[]{multicol}
\usepackage{wrapfig}
\usepackage{authblk}

\usepackage{floatflt}

\title{\large Complete list of the ASTRO-H Science Working Group}
\date{\vspace{-0.5cm}}
\newcommand{\MakeWhitePaperTitle}{
	\begin{center}
		\begin{figure}
			\vspace{1cm}
			\begin{center}
				\includegraphics[width=0.2\hsize]{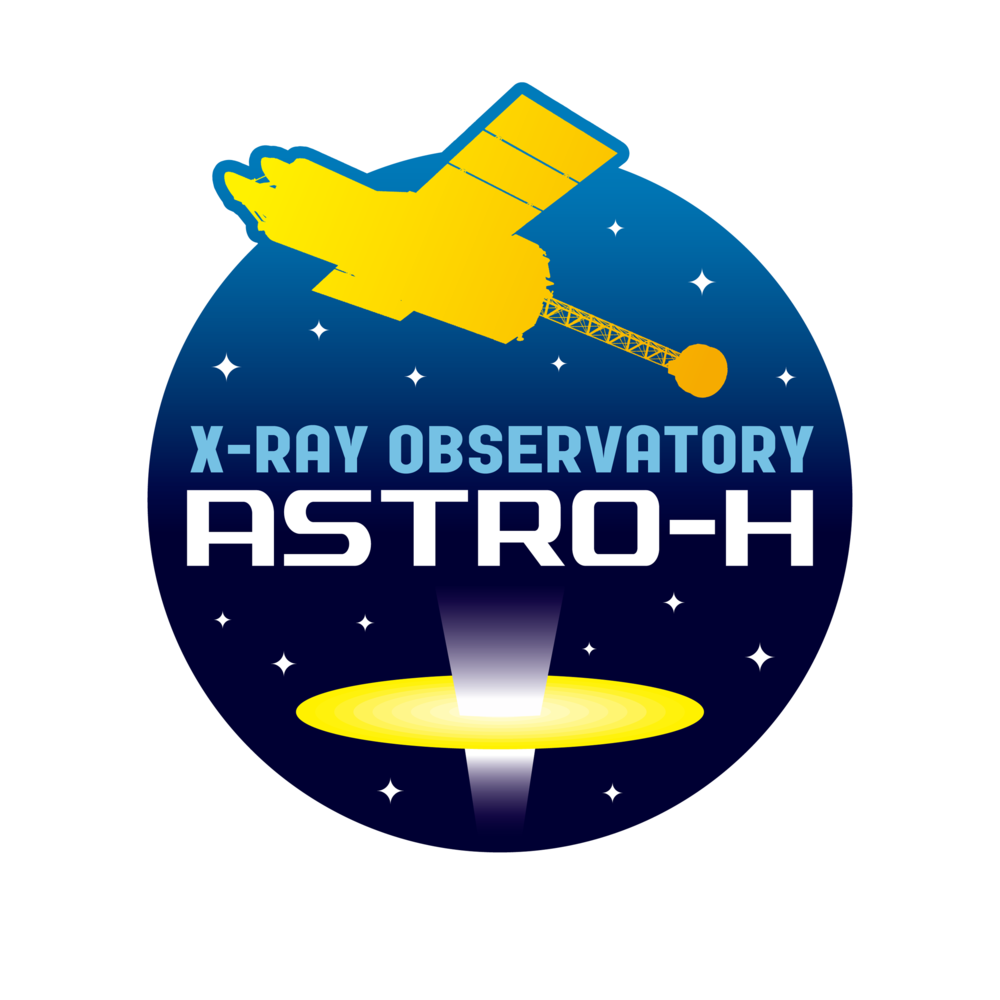}
			\end{center}
		\end{figure}
		\vspace{1cm}
		{\LARGE
		ASTRO-H Space X-ray Observatory\\
		White Paper\\
		}
		\vspace{5mm}
		{\large
		\WhitePaperTitle\\
		}
		\vspace{1cm}
		{
		\WhitePaperAuthors\\
		on behalf of the ASTRO-H Science Working Group
		}
	\end{center}
}

\usepackage{authblk}
\author[a]{Tadayuki~Takahashi}
\author[a]{Kazuhisa~Mitsuda}
\author[b]{Richard~Kelley}
\author[c]{Felix~Aharonian}
\author[d]{Hiroki~Akamatsu}
\author[e]{Fumie~Akimoto}
\author[f]{Steve~Allen}
\author[g]{Naohisa~Anabuki}
\author[b]{Lorella~Angelini}
\author[h]{Keith~Arnaud}
% ...
\affil[a]{Institute of Space and Astronautical Science (ISAS), Japan Aerospace Exploration Agency (JAXA), Kanagawa 252-5210, Japan}
\affil[b]{NASA/Goddard Space Flight Center, MD 20771, USA}
\affil[c]{Astronomy and Astrophysics Section, Dublin Institute for Advanced Studies, Dublin 2, Ireland}
\affil[d]{SRON Netherlands Institute for Space Research, Utrecht, The Netherlands}
\affil[e]{Department of Physics, Nagoya University, Aichi 338-8570, Japan}
\affil[f]{Kavli Institute for Particle Astrophysics and Cosmology, Stanford University, CA 94305, USA}
\affil[g]{Department of Earth and Space Science, Osaka University, Osaka 560-0043, Japan}
\affil[h]{Department of Astronomy, University of Maryland, MD 20742, USA}
% ...

\begin{document}

%---------------------------------------------
% title
%---------------------------------------------
\newcommand{\WhitePaperTitle}{
	High Resolution Spectroscopy of Interstellar and \\
	Circumgalactic Gas in the Milky Way and Other Galaxies
}
\newcommand{\WhitePaperAuthors}{
F.~Paerels~(Columbia~University),
N.~Yamasaki~(JAXA),
N.~Anabuki~(Osaka~University),
E.~Costantini~(SRON),
C.~de~Vries~(SRON),
R.~Fujimoto~(Kanazawa~University),
A.~Hornschemeier~(NASA/GSFC),
R.~Iizuka~(JAXA),
C.~Kilbourne~(NASA/GSFC),
S.~Konami~(Tokyo~Metropolitan~University),
S.~LaMassa~(Yale~University),
M.~Loewenstein~(NASA/GSFC~\&~University~of~Maryland),
D.~McCammon~(University~of~Wisconsin),
K.~Matsushita~(Tokyo~University~of~Science),
B.~McNamara~(University~of~Waterloo),
I.~Mitsuishi~(Nagoya~University),
R.~Nagino~(Osaka~University),
T.~Nakagawa~(JAXA),
S.~Porter~(NASA/GSFC),
K.~Sakai~(JAXA),
R.~K.~Smith~(SAO),
Y.~Takei~(JAXA),
T.~Tsuru~(Kyoto~University),
H.~Uchiyama~(Shizuoka~University),
H.~Yamaguchi~(NASA/GSFC~\&~University~of~Maryland),
and S.~Yamauchi~(Nara~Women's~University)
}
\MakeWhitePaperTitle

\begin{abstract}
We describe the potential of high resolution imaging spectroscopy with the SXS on {\it ASTRO-H} to
advance our understanding of the interstellar- and circumgalactic media of our own Galaxy, and other galaxies. Topics to be addressed range from absorption spectroscopy of dust in the Galactic interstellar medium, to observations to constrain the total mass-, metal-, and energy flow out of starburst galaxies.
\end{abstract}
\clearpage

%---------------------------------------------
% member list
%---------------------------------------------
\maketitle
\clearpage

%---------------------------------------------
% TOC
%---------------------------------------------
\tableofcontents
\clearpage

%%%%%%%%%%%%%%%%%%%%%%%%%%%%%%%%%%%%%%%%%%%%%%%%%%
\section{Introduction}
%%%%%%%%%%%%%%%%%%%%%%%%%%%%%%%%%%%%%%%%%%%%%%%%%%%

The microcalorimeter spectrometer on {\it ASTRO-H} (Soft X-ray Spectrometer; SXS) will be the first high-resolution imaging X-ray spectrometer in space. We will be able to perform sensitive emission- and absorption spectroscopy on the diffuse interstellar and circumgalactic media both in our own, and in external galaxies. We describe the potential for addressing a range of topics, from studying the physical chemistry of Galactic dust, to the `Missing Metals' problem. We briefly introduce the main topics below.

There is a sizable body of high resolution absorption line spectroscopy of the Galactic interstellar medium (ISM) and halo/Intragroup medium, obtained with the diffraction grating spectrometers on {\it Chandra} and {\it XMM-Newton} along lines of sight towards bright point sources. These spectrometers cannot perform spectroscopy of extended sources. If we complement the grating spectroscopy with measurements of emission line intensities along the same, or close-by lines of sight with the {\it ASTRO-H} SXS, we can determine the average density and length scale along each sight line \citep[e.g.][]{yao2007,yoshino09}. The obvious improvement over data obtained with CCD spectrometers will be in the ability of the  SXS to resolve the Fe L complex, and hence to obtain an unambiguous and improved measurement of the electron temperature(s) of the Circumgalactic gas, as well as the Fe/O abundance ratio. 

The SXS will be the first astrophysical high-resolution spectrometer (virtually) without Si in the detectors. 
It also has by far the highest sensitivity of any spectrometer at Fe K. We will be able to turn both features to advantage to perform the first sensitive study of X-ray absorption by solid material (dust) in the ISM, at the Si and Fe K edges, using the XAFS technique.

Obviously, the SXS has the potential to provide us with emission line spectroscopy of diffuse gas (and unresolved point source populations) in external galaxies. Two target classes are discussed below,  starburst galaxies and elliptical galaxies. We can address the physical state and chemical composition of the ISM, and determine outflow velocities from starbursts, and the implied metal mass flux into the intergalactic medium.

With the SXS, we will also have the exciting prospect of extending the study of the circumgalactic medium to other galaxies. Due to source confusion and a lack of sensitivity, absorption spectroscopy of point sources in other galaxies has not been possible with the diffraction grating spectrometers. With the SXS, this will be possible, using the brightest Ultra-Luminous X-ray sources observed in certain galaxies. In deep exposures, we will also be able to perform the complementary emission line spectroscopy on galactic halo's with high surface brightness and large angular extent. We will also exploit the few coincidences between galaxies and bright background AGN for absorption spectroscopy.

%%%%%%%%%%%%%%%%%%%%%%%%%%%%%%%%%%%%%%%%%%%%%%%%%%%
\section{Galactic Interstellar Dust Absorption}
%%%%%%%%%%%%%%%%%%%%%%%%%%%%%%%%%%%%%%%%%%%%%%%%%%%

\subsection{Background and Previous Studies}

The interstellar matter bears the signature of the star formation and stellar evolution in our Galaxy. Dust 
is efficiently produced by late-type stars and supernova explosions. In turn, dust and gas provide 
the reservoir  of matter for the formation of new stars and planets. Dust regulates the thermal balance in star formation regions at all red shifts. After decades of multi-wavelength (from radio to far-ultraviolet)
studies, the broad characteristics of interstellar dust have been established \citep[e.g.][]{draine09}. There are however
important open issues, for instance the nature of the iron content of dust grains, as only a fraction of it is
included in silicates \citep{whittet03}. Most of the astrophysical dust material appears to be in an amorphous phase, where the long-range crystalline structure has been broken up, either by the grain growth process, or by subsequent processing in the ISM; however, the degree of disorder is unknown. X-ray spectroscopy can address these issues directly.
 
It has become clear that advanced X-ray instruments are a powerful tool to characterize both the 
gas and dust phase of the ISM in our Galaxy. 
Dust both absorbs and scatters the X-ray radiation. 
 The background light of bright X-ray emitters, as X-ray binaries are, allows us to study dust by both mechanisms. 
 Absorption can be studied through high-signal-to-noise, high resolution absorption spectra. 
 Here, the sensitivity to the short-range order or `micro-crystalline' structure of the dust material arises from quantum interference of outgoing and scattered photoelectron wave functions, which gives rise to a unique `X-ray Absorption Fine Structure' (XAFS) in the photoelectric absorption spectrum near absorption edges \citep[e.g.][and references therein]{deVries2009,lee09}.
 Scattered light, 
 which results in a halo of diffuse emission around the source, is studied thanks to the unique feature of 
 imaging coupled with spectral resolution of the SXS.  
 X-ray binaries are mostly located within the Galactic 
 plane, allowing to study the ISM in a variety of environments, with different dust and gas enrichment history. 
 The information on the ISM therefore is embedded in every observation carried out on compact objects.
The SXS instrument will access a spectral region up to now largely unexplored, in particular the 
Mg K, Si K and Fe K edge region, located between 1.3 and 7.1 keV. These 
are important to investigate the nature of the silicates in our Galaxy and in particular 
the  iron inclusion in them, as in the X-ray band, unlike in other wavelength ranges, as many as two 
strong neutral iron features may be observed. The Fe K feature is an important metallicity indicator 
in the densest regions, like e.g. the Galactic Center, 
where absorption is such as to obliterate all other lower energy features. 

The SXS will be able to access routinely the Si K edge, as this becomes a deep absorption feature for hydrogen column
densities which are commonly measured towards X-ray binaries ($N_{\rm H}\sim1-5\times10^{22}$\,cm$^{-2}$). 

\subsection{Prospects and Strategy}
For the Si K edge, a number of
persistently high-flux X-ray binaries can be observed. Indeed many of the brightest X-ray binaries lie behind a relatively high column density ($N_{\rm H}\sim1-3\times10^{22}$\,cm$^{-2}$). The best targets for this kind of studies are Low Mass X-ray Binaries (LMXB), which display generally a featureless continuum. Therefore only absorption features (lines and edges) from dust and both neutral and possibly ionized gas are visible. In other classes of bright sources, like the High-Mass X-ray binaries or Supernova remnants is nearly impossible to study dust features from the ISM, because of the complex emission spectrum displayed by these sources.
At the Si K edge, the resolution of SXS will be similar to the one of the {\it Chandra} HETG, but with higher effective area. Therefore we expect to resolve with high accuracy different possible chemical compositions of the dust (Figure~\ref{f:si}). 
There is only a handful of sources which naturally lie in very dense environments ($N_{\rm H}>7\times10^{22}$\,cm$^{-2}$), which would maximize the depth of the Fe K edge. In order to reach an acceptable signal to noise ratio, these sources must be in outburst. This is a crucial issue, as it will be the signal to noise level, rather than the resolution, that determines the feasibility of studying dust in the Fe K region.
Several transients have been detected in the past behind a high column density medium. Table 1 lists five favorable binaries; the Fe K  targets have to be observed during outburst or a high level flux. For this, a triggered observation will be needed.

\subsection{Target and Feasibility}

\begin{figure}
\begin{center}
\includegraphics[width=7cm,angle=90,bb=0 0 426 579]{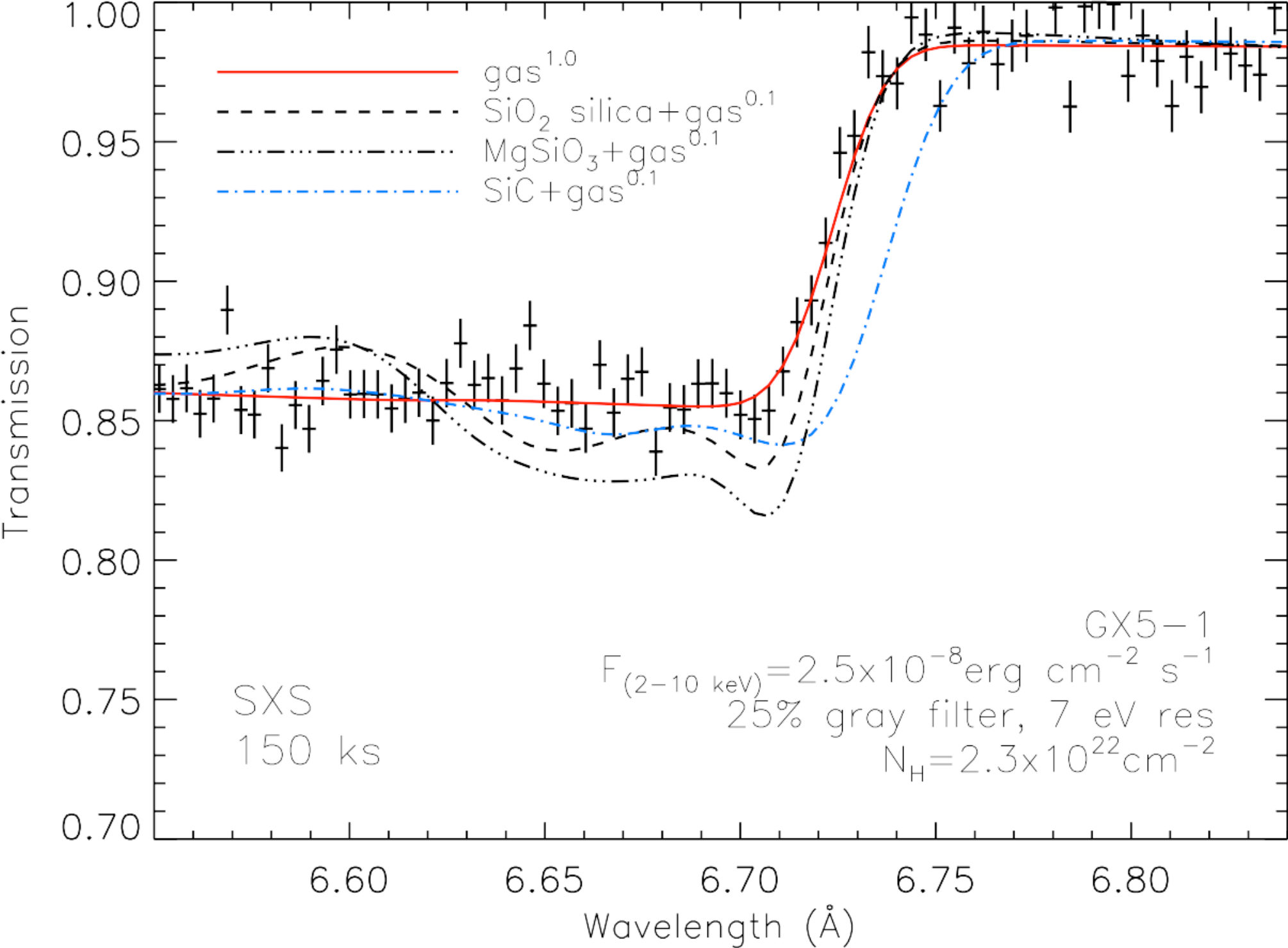}

\end{center}
\caption{\label{f:si} 150\,ks SXS simulation for a typical bright X-ray binary (here GX~5-1) in the Si\,K region, with a 
\ $2-10$ keV flux of $F_{\rm X} = 2.5 \times 10^{-8}$ erg cm$^{-2}$ sec$^{-1}$.
The 25\% transmission neutral density filter has been inserted to limit the count rate, and an energy resolution of 7 eV has been assumed.
Different dust components plus gas (10\% of the total) are compared to absorption by Si in the gas phase only. The exposure time is 150 ks, 
}
\end{figure}

\begin{figure}
\begin{center}
\includegraphics[width=7cm,height=10cm,angle=90, bb=0 0 425 595]{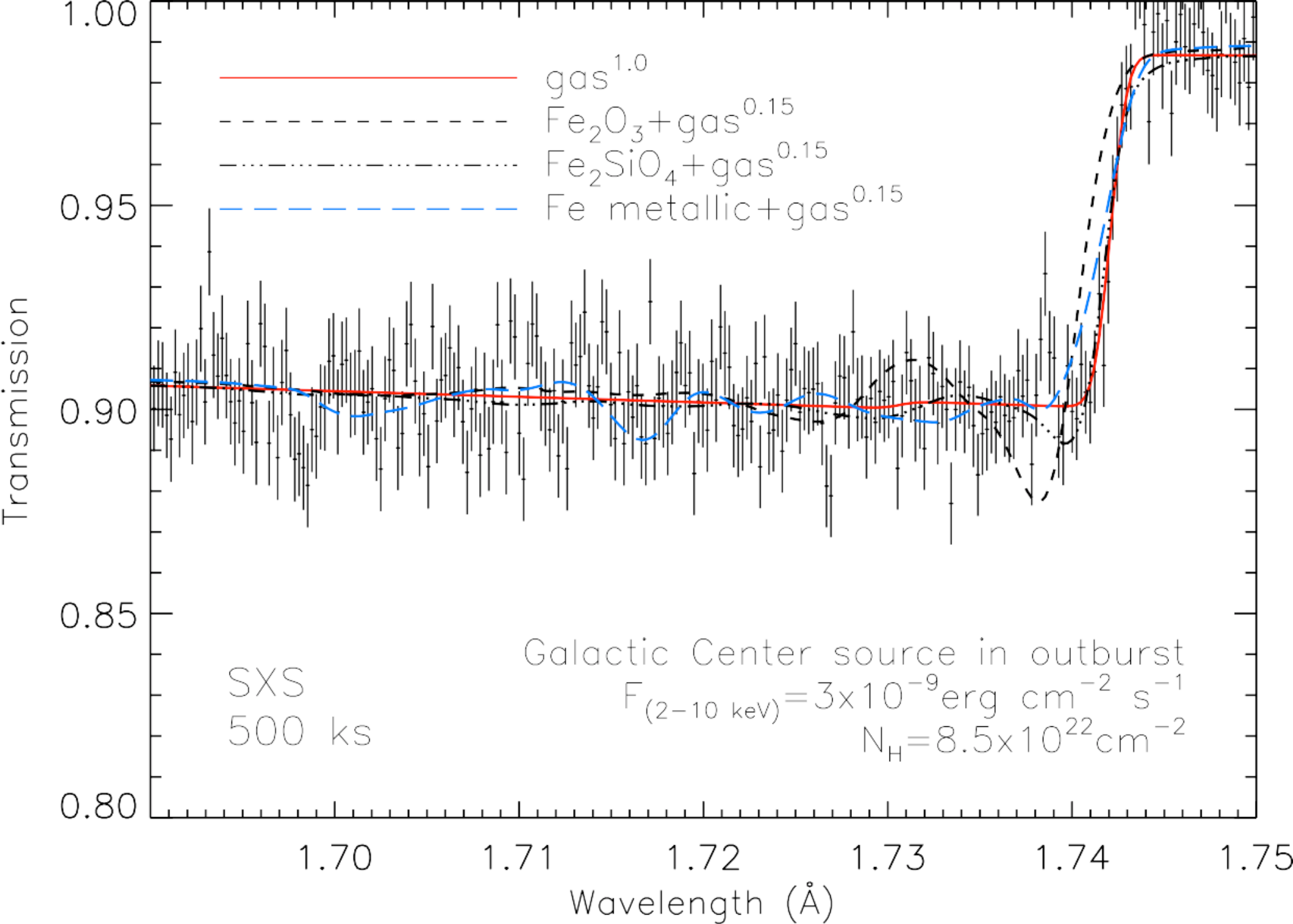}
\end{center}
\caption{\label{f:fe} 500\,ks SXS simulation for an outbursting X-ray source near the Galactic center (here
SAXJ1747.0-2853) in the Fe\,K region, with an assumed\ $2-10$ keV flux of  $F_{\rm X} = 3 \times 10^{-9}$ erg cm$^{-2}$ sec$^{-1}$. The assumed energy resolution is 7 eV. 
Different dust components plus gas (15\% of the total) are compared to absorption by Fe in the gas phase only.
}
\end{figure}

In Figure~\ref{f:si} an example of different edge shapes due to different dust compounds is shown for the
Si~K edge. In Figure~\ref{f:fe} the Fe~K edge is shown for a source with high-$N{\rm _H}$ caught in a high flux
state \citep[e.g.][]{intzand04}. The dust detection here depends on the nature of the dust, as some compounds show little deviation from a `normal' ({\it i.e.} atomic)
photoelectric edge (dust absorption profiles from \citet{lee09}). 

For the Si K simulations, we
used a mixture of gas (10\%) and two common dust 
compounds\footnote{Source of dust profiles: http://www.sron.nl/files/HEA/XRAY2010/talks/3/lee.pdf}, SiO$_2$ and MgSiO$_3$, 
equally distributed in order to reach a proto-solar abundance for Si \citep{lodders09}. 
We estimated the exposure time imposing the requirement that the determination 
of the abundance of Si in gas form (i.e. a measure of depletion) be significant to 5$\sigma$. 
Detection of dust in the Fe~K edge is difficult. Here we considered a mixture of gas (15\%) plus one dominant dust
component (Fe$_2$SiO$_4$). The total Fe abundance has been set to proto-solar. The exposure time has been calculated
imposing the requirement that the determination of the abundance of Fe in gas form (i.e. a measure of depletion) to be significant at the 5$\sigma$ level.

\begin{table}
\begin{center}
\caption{List of sources suitable for dust detection by SXS}
\begin{tabular}{lllll}
\hline
source & $N_{\rm H}$ & Flux 2--10 keV & edge & exposure\\
& $10^{22}$cm\,$^{-2}$ & $10^{-9}$\,erg\,cm$^{-2}$\,s$^{-1}$ & & ks\\
\hline
GX5--1 & 3 & 25 & SiK & 150\\
%GX13+1 & 1--3 & 9 & SiK & (300)\\
GX340+0 & 5 & 9.5 & SiK & (300)\\
\hline
SAX1747.0-2853 & 8.5 & 4 & FeK & 300\\
CirX-1 & 9 & 8 & FeK & (250)\\

\hline
\hline
\end{tabular}
\\
\end{center}

{\it notes: exposure times calculated in order to have a 5$\sigma$ significance on the  detection of absorption by Si or Fe in the gas phase. For
the Si-sources absorption by gas plus two dust components was assumed.
For the Fe-sources only gas and one dust component have been assumed. For Fe, 
the actual dust detection strongly depends on the type of dust. These sources must be observed in a high-flux state to study iron.\\
The exposures for sources not proposed as PV targets are reported in brackets.
}
\end{table}

%%%%%%%%%%%%%%%%%%%%%%%%%%%%%%%%%%%%%%%%%%%%%%%%%%%
\section{Hot Interstellar and Circumgalactic Gas}
%%%%%%%%%%%%%%%%%%%%%%%%%%%%%%%%%%%%%%%%%%%%%%%%%%%

\subsection{Background and Previous Studies}

Studies of hot gas in the circum--Galactic environment, based on X-ray absorption- and emission line spectroscopy, 
have provided evidence for a reservoir of $kT_e \sim 0.2$ keV gas. 
As a typical value, the emission is as dim as a few to several photons cm$^{-2}$ sec$^{-1}$sr$^{-1}$ in the strongest Oxygen K-shell lines. 
Combining absorption line columns and line emission measures allows one to estimate characteristic gas densities and path lengths. 
Interpretations of the existing data vary (see for instance \cite{yao2012}). 
{\it Suzaku}'s clean low-energy CCD response has allowed a study of emission line intensities in several directions \citep{yoshino09}. 
This has revealed revealed a constant contribution to the surface brightness of O${\rm VII}$ line emission, observed in several directions \citep{yoshino09}.
This component is present even in shadowing observation towards nearby molecular clouds,  {\it e.g.} MBM20, at a distance of $\sim$ 150 pc.
This indicates that some part of the soft X-ray background comes from 
our  vicinity. Two major candidates are the so-called "Local Hot Bubble", which  might be a local Supernova remnant with hot plasma, 
and Solar wind charge-exchange (SWCX) at the edge of heliosphere \citep{Gupta2009}. 
As {\it Suzaku} observed a temporal change of the O${\rm VII}$ line surface brightness toward the Lockman Hole, 
the contribution from SWCX is not negligible \citep{Yoshitake2013}. This topic will be discussed in 
another {\it ASTRO-H} White Paper (17).

The higher resolution of the SXS in principle will allow more precise constraints to be placed on the properties of the diffuse gas. 
For instance, resolving the Fe L complex at 800 eV can produce an unambiguous measurement of the electron temperature. 
Combined with the O emission lines, we can measure the Fe/O ratio; there is evidence that this ratio varies across the sky. 
Also, the interpretation of the measured O VII absorption column densities hinges critically on the assumed ionization balance. 
At gas temperatures around $kT\sim 0.2$  keV, that correction varies strongly with 
temperature. 

\subsection{Prospects}
\begin{figure}
\begin{center}
\includegraphics[width=10cm,bb=0 0 792 612]{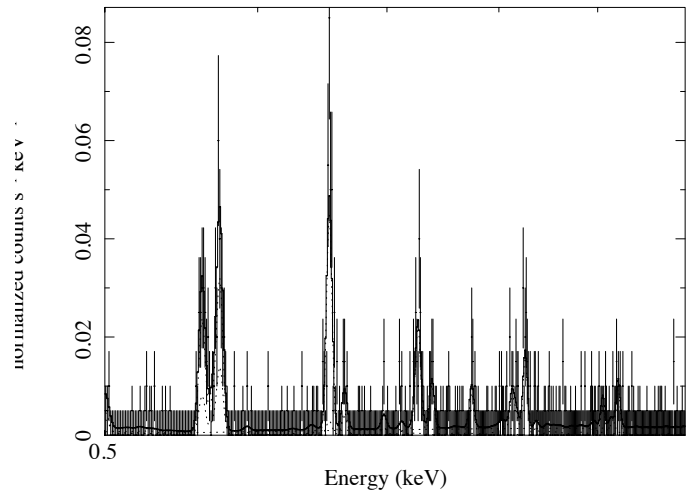}
\end{center}
\caption{\label{nep-200k} 200\,ks SXS simulation for a field in the North Ecliptic Pole region. The fore-and backgrounds have been modeled as described in \cite{yoshino09}; the parameters for the circum--Galactic emission are $kT_e = 0.25$ keV, emission measure $9.9 \times 10^{14}$  cm$^{-5}$ sr$^{-1}$ (gas in collisional ionization equilibrium). The four strongest emission features are (low to high energy): 
the O VII $n=1-2$ `triplet' (resonance and forbidden transitions resolved), O VIII Ly$\alpha$, 
Fe XVII $2p-3s$ ($\lambda\lambda 17.05, 17.10$ \AA; 725, 727 eV), Fe XVII $2p-3d$ ($\lambda 15.014$ \AA, 825 eV).  Faint O VII and VIII $n=1-3$ transitions are also visible.
}
\end{figure}

We can pursue the measurement of the diffuse Galactic emission line 
spectrum with pointed observations, in addition to using archival data of 
near-blank fields (weak point source targets) and possibly accumulated 
slew data. The small field of view of the SXS requires deep observations 
to obtain significant emission line detections. Based on the {\it Suzaku} data 
gathered in \cite{yoshino09}, we estimate that at least 200 ksec exposure 
on a single field is required. Figure~\ref{nep-200k} shows a simulated 
spectrum for 200 ksec exposure on a field near the North Ecliptic Pole, 
which will determine the O Ly $\gamma$ line intensity with 
an accuracy of 30\%. A single pointing to obtain 
an accuracy of 10\% will require 
an exposure of order a Msec for SXS. This will be a challenging 
observation, since the uncertainty in the emission model is large, and this 
observation therefore appears less suitable for the PV phase schedule.
Archival study in a later stage of the mission based on a compilation of 
blank sky data, on the other hand, may allow us to obtain unique 
information on the circumgalactic gas. The SWCX component may be 
identifiable through time variability correlated with changes in the Solar 
wind, or Coronal Mass Ejection Events, by tracking intensity variations in 
the O Ly $\gamma$ line (which is relatively bright in recombining plasmas 
as opposed to collisional plasmas).

%%%%%%%%%%%%%%%%%%%%%%%%%%%%%%%%%%%%%%%%%%%%%%%%%%%
\section{Starburst Galaxies and the Missing Metals}
%%%%%%%%%%%%%%%%%%%%%%%%%%%%%%%%%%%%%%%%%%%%%%%%%%%

\subsection{Background and Previous Studies}

Galaxies and the Intergalactic Medium (IGM) are connected by flows of matter and energy. Both accretion onto galaxies and outflows from galaxies or their central black holes link galaxies to the IGM in what has been termed "Cosmic Feedback". To obtain a deeper physical understanding of both galaxy formation and evolution and the IGM requires that we better understand the physical processes that link them. 
  
Mechanical feedback (stellar winds and SNe) is also the primary physical mechanism creating the hot phases of the interstellar and circumgalactic media (ISM and CGM) in star-forming galaxies (spiral, irregular and merging galaxies). The plasmas making up the hot phases of the ISM/CGM have temperatures in the range $T =10^6 - 10^8$~K and predominantly emit and absorb photons in the X-ray energy band from $E=0.1 - 10$~keV. Line emission (from highly-ionized ions of the astrophysically important elements O, Ne, Mg, Si, S, and Fe) dominates the total emissivity of plasmas with temperatures $T<10^7$ K, and at higher temperatures Ar, Ca and in particular Fe also produce strong lines.  For a longer description of the outflow of metals and energy into the IGM, we refer the reader to the recent Astro2010 decadal white paper by David Strickland (paper 89, http://sites.nationalacademies.org/BPA/BPA\_050603\#galaxies). 

The hot phase of the ISM dominates the energetics of the ISM and strongly influences its phase structure \citep{efstathiou00}. X-ray observations are a natural and powerful probe of the composition and thermodynamic state of hot phases of the ISM/CGM in and around galaxies, and thus are also a powerful tool for exploring the physics of feedback, as well as completing the `metal census', and following the 
enrichment of the IGM.

Recent results from {\it XMM-Newton} and {\it Suzaku} have revealed much about the ISM in starburst galaxies.
Most recently, \cite{mitsuishi13} conducted a through study of the nearby starburst galaxy NGC 253, 
the ISM of which is well-modeled by a two-temperature plasma ($kT \sim$ 0.2 keV and $kT \sim$ 0.6 keV) 
outside the nuclear region where the intense starburst activity occurs.
On the other hand, \cite{mitsuishi11}  shows the existence of hotter plasma ($>$ 1 keV) 
possibly originating from the starburst activity in the nuclear region of NGC 253.
Using detailed abundance measurements in three different regions from the central part of NGC 253 
to the halo ($\sim$10 kpc away from the center) within the galaxy they found all three regions were consistent 
with Type II supernova ejecta, indicating that the starburst activity in the central region 
provides metals toward the halo through a galactic-scale starburst-driven outflow. 
The outflow velocity was constrained to be $>$ 100 km s$^{-1}$  from the X-ray data 
which may be compared with the $\sim$220 km s$^{-1}$ escape velocity of NGC 253. 
Note that outflow velocities in the cooler entrained material (molecular cloud, cooler gas and dust) 
have been measured as well 
(50--400 km s$^{-1}$ in CO, 100--300 km s$^{-1}$ in Hα, and 300--2000 km s$^{-1}$ in the far-infrared (dust outflow); 
\citet{Balatto2013}; \citet{westmoquette11}; \citet{kaneda09} respectively).

Similar results were obtained with {\it Suzaku} on M82, finding via detailed abundance measurements that starburst activity enriches the M82 outflow through SNcc metal ejection  \citep{Konami2011}. 
{\it Suzaku} studies of farther-out regions in M82, including the M82 Cap located 11$^{\prime}$ (11.6~kpc) away from the nucleus have demonstrated metals consistent with Type II supernovae even at these great distances \citep{tsuru07}.   

While significant progress has been made in measuring abundances in hot gas, 
it should be noted that SXS with excellent energy resolution provides us with new possibilities  
to measure the outflow velocity of X-ray emitting hot gas directly.
The starburst phenomenon itself is considered to be universal in the history of the Universe \citep[e.g.][]{Pettini2001}
and a normal event in the life of a galaxy \citep[e.g.][]{McQuinn2010}, and plays a key role in transporting material and energy 
into intergalactic space.
%Thus, to "capture" an outflowing motion of material associated with a starburst activity will allow us to establish 
%a scenario of evolution of the Universe both from chemical and dynamical points of view.
Although X-ray emitting hot gas heated by successive supernovae is a direct tracer of the outflowing material,  
to date all existing observational velocity measurements are of entrained cooler material, 
e.g., molecular clouds and warm neutral and ionized gas, with outflow velocities in the range 200--1000 km s$^{-1}$ 
measured using radio/optical/UV emission and/or absorption lines \citep[e.g.][]{shapley03,rupke05,martin05}.
Therefore, detecting the outflow velocity of the hot gas is the last piece in the verification of the starburst-driven outflow scenario.

\subsection{Prospects \& Strategy}

The SXS will revolutionize studies of abundances and temperatures in 
starburst galaxies, allowing important conclusions regarding the physical 
state of the hot gas. {\it Suzaku} spectroscopic studies have already 
demonstrated how well this works 
\citep[e.g.][]{tsuru07,Konami2011,mitsuishi13} so we can naturally 
continue and extend this work into more precisely determining the state of 
the ISM and therefore modeling the flow of energy and metals to the IGM.    
Thus the prospects are excellent and the tools/machinery/expertise is well 
in hand to conduct thorough studies.

The ultimate smoking gun would be the direct measurement of the velocities of the outflowing gas and demonstrate that they exceed the escape velocities of the galaxies.   With the SXS, we will in principle be able to see Doppler {\it shifts} from outflowing gas (in practice, Doppler {\it broadening} will probably be too hard to detect). With sufficient signal-to-noise, one can in principle detect shifts of order a fraction (a few tenths) of the energy resolution, or maybe $\Delta E > 1$ eV, which corresponds to $v > 300$ km s$^{-1}$ at 1 keV and $v > 50$ km s$^{-1}$ at 6 keV. However, this assumes a perfect knowledge of the gain. This will be challenging if the gain uncertainty for the SXS is at the 2 eV requirement.  We describe the prospects for velocity measurements below in ``Beyond Feasibility."

\subsection{Targets \& Feasibility}

\begin{figure}
\begin{center}
\includegraphics[angle=0,scale=0.50]{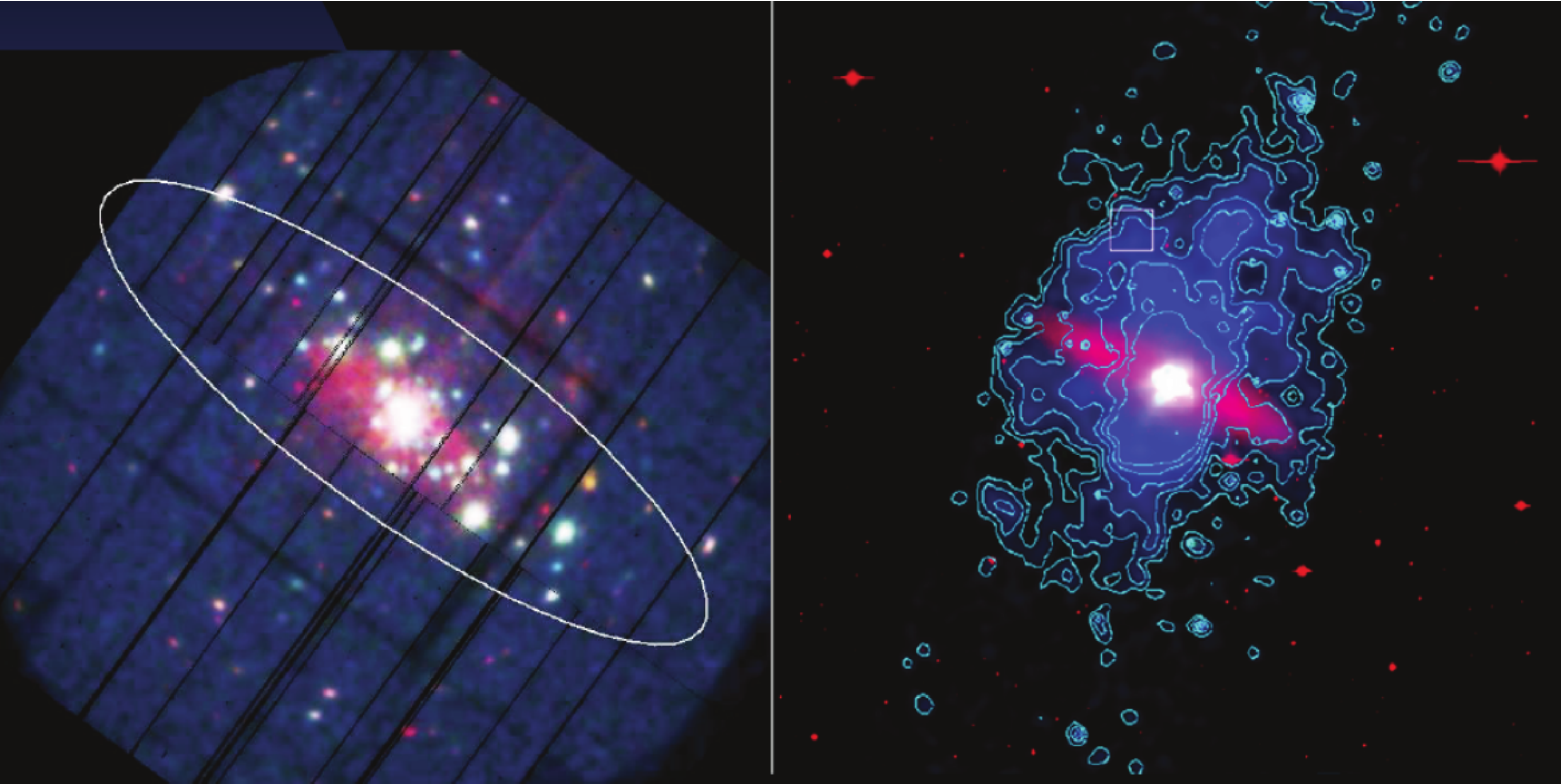}
\caption{\label{starbursts_im} NGC 253 ({\it left panel}) and M82 ({\it right panel}) are the two nearest and X-ray brightest starburst galaxies, which make them the best candidates for detailed spectroscopic studies of starburst outflows.  Shown are {\it XMM-Newton} and {\it Chandra} images of NGC 253 and M82 respectively.  The large ellipse on the NGC 253 image is the D25 extent of the disk ($25^{\prime} \times 7^{\prime}$).  M82 is more compact, the small box is roughly $15^{\prime \prime}$ on a side.}
\end{center}
\end{figure}

There are approximately 30 starburst galaxies in the local universe (within $d \approx 100$ Mpc) with star formation properties and X-ray constraints that make them candidate targets for study with a high-resolution X-ray spectrometer (Strickland et al. 2009).   However, of these, only two are sufficiently bright for consideration by {\it ASTRO-H}, and the exposures may be long enough that one should ultimately be chosen.  These are NGC 253 ($d \approx2.6$ Mpc) and M82 ($d \approx 3.6$) Mpc.  These two galaxies differ in a few key ways.  NGC 253 is very large on the sky with a D25 extent of $25^{\prime} \times 7^{\prime}$, making it possible to select different regions within the galaxy for study.  M82 is more compact {\it and} more luminous.  
Figure~\ref{starbursts_im} shows X-ray images of both galaxies.

\begin{figure}
\begin{center}
\includegraphics[width=0.45\textwidth, bb=0 0 1024 768]{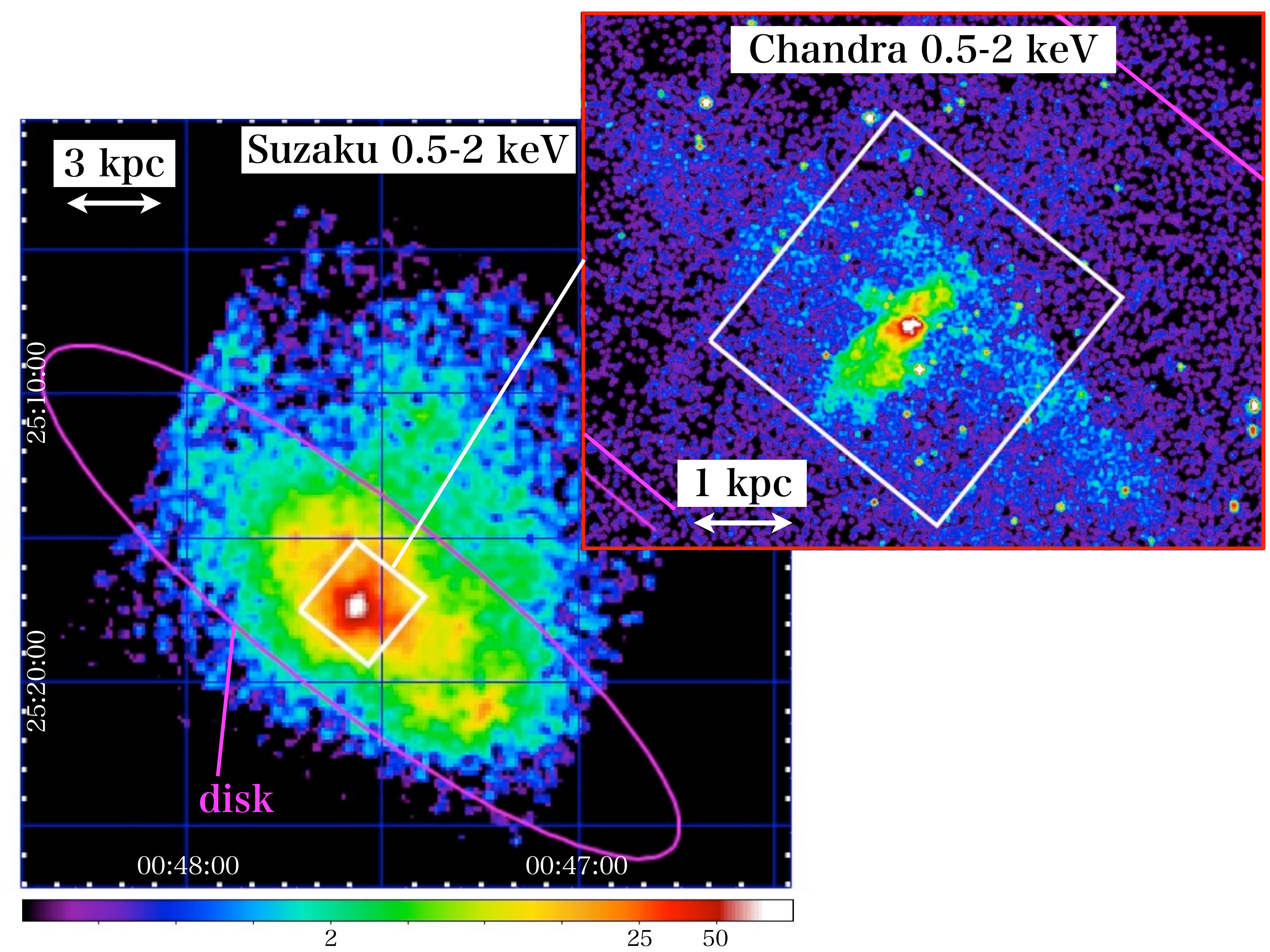}
\includegraphics[width=0.45\textwidth, bb=0 0 1024 768]{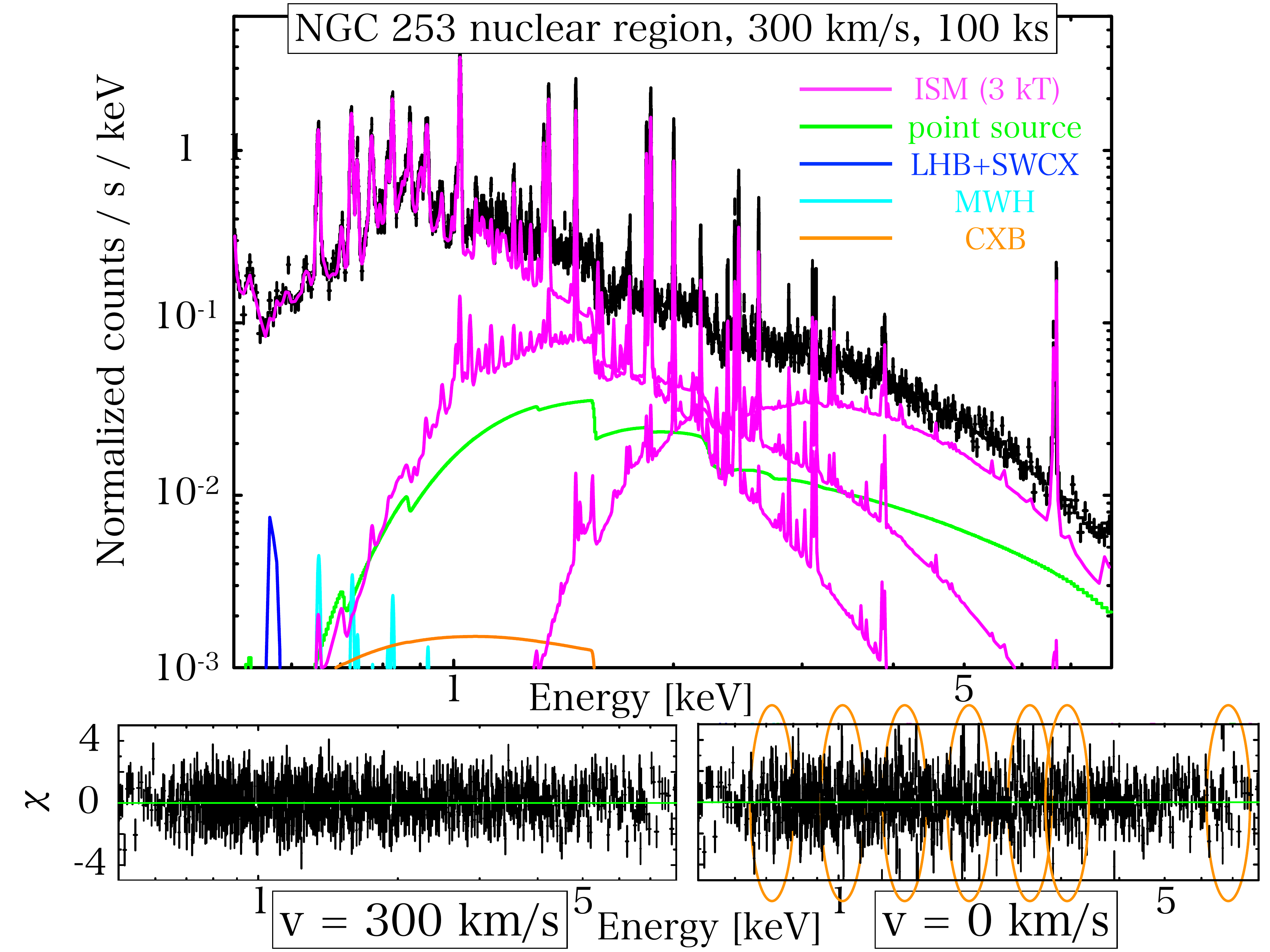}
\caption{Left: {\it Suzaku} and {\it Chandra} images of NGC 253 in 0.5-2.0 keV. White square and magenta ellipse indicate the proposed region and the optical disk. Right top: an expected SXS spectrum of the proposed region with an exposure time of 100 ks fitted with three-temperature ISM (magenta), point sources (green), and X-ray background (blue, cyan and orange) models. Blue is the contribution from the Local Hot Bubble (LHB) and Solar Wind Charge Exchange (SWCX)emission;  cyan is the emission from the Galactic (Milky Way) Halo (MWH), and orange is the contribution from the unresolved Cosmic X-ray Background (CXB).
Right bottom: residuals between the models and the data with outflow velocities of 300 km s$^{-1}$ and 0 km s$^{-1}$. Medium energy resolution of 7 eV is assumed. The orange ellipses emphasize post-fit residuals at the strongest emission lines if a spectrum simulated with a velocity shift of 300 km sec$^{-1}$ is fit with a model with a zero velocity shift.}
\label{fig:NGC253}
\end{center}
\end{figure}

%%NGC253
For instance in Mitsuishi et al. (2011) and Mitsuishi et al. (2013) there were four distinct regions studied in NGC 253 
and we would likely suggest observing the nucleus to measure the outflow velocity of the observed hot gas. 
Its sufficient brightness ($\sim$10$^{-12}$ erg/s/cm$^{2}$ in 0.5--2~keV) in the SXS Field of View (FOV)`, ISM-dominant situation, large apparent size ($27' \times 7'$ in optical disk), 
substantial multi-wavelength studies, and wide opening angle ($\sim 60^{\circ}$, \cite{westmoquette11}) make the galaxy the best target in this study. 
Figure \ref{fig:NGC253} shows {\it Suzaku} and {\it Chandra} images of the proposed region and an expected SXS spectrum 
with an exposure time of 100 ks and an outflow velocity of 300 km s$^{-1}$.
Here, 300~km s$^{-1}$ is assumed as a reasonable value considering other wavelength studies \citep[e.g.][]{Balatto2013,westmoquette11,kaneda09}
and an assumption that hot gas motion is faster than other components.
Thanks to the excellent energy resolution of SXS with good photon statistics, typical statistical errors on temperatures and metal abundances are 10--40\% 
at the 90\% significance level, which enable us to investigate detailed plasma diagnostics. 
The outflow velocity is also tightly constrained, on the order of 50 km s$^{-1}$, even with an exposure time of 50 ks within 5\% accuracy assuming ideal gain calibration.
However, considering the uncertainty on the gain scale (requirement: 2 eV, goal: 1 eV), emission lines at higher energies such as He-like Ar (3.14 keV), Ca (3.90 keV), 
and Fe (6.70 keV) should be utilized to distinguish an energy shift derived from the outflow from the gain uncertainty.
Thus, to obtain sufficient photon numbers for these lines, 100 ks observation is required to determine the outflow velocity within ~30\% accuracy based on a fit to the 3--8 keV band.
Table \ref{table:NGC253feasibility} summarizes the measurement accuracies as a function of exposure time. 

\begin{table}
\renewcommand{\baselinestretch}{1.1}\selectfont
\small{
  \caption{NGC 253: Expected Exposure Time$^{\ast}$.}
  \vspace{-0.3cm}
    \label{table:NGC253feasibility}
  \begin{center}
    \begin{tabular}{cccccc}
\hline\hline
        & Exposure [ks]                              &    50                         &        100             &     200               &     300    \\ \hline
        & Accuracy$^{\dagger}$ [\%]        &    $<$5                     &  $<$5                 &     $<$5            &    $<$5  \\ \cline{3-6}
        &                                                                                         &    -$^{\ddagger}$  &  $\sim$30        &     $\sim$30    &    $\sim$20  \\
\hline
    \end{tabular}
\begin{flushleft}
\renewcommand{\baselinestretch}{1.2}\selectfont
\footnotesize{
  \vspace{-0.2cm}
\hspace{2.3cm}$^{\ast}$Estimated exposure time for an outflow velocity of 300 km s$^{-1}$. \\  
\hspace{2.3cm}$^{\dagger}$Resultant redshift does not include the input value within the 90\% confidence level. \\
\hspace{2.3cm}$^{\ddagger}$Measurement accuracy for the fit to 0.5--8 keV (top) and 3-8 keV (bottom).
}
\end{flushleft}
  \end{center}
}
\end{table}
\renewcommand{\baselinestretch}{1.0}\selectfont

%%M82
The central region of M82 is very bright in X-rays, but it consists not only of diffuse hot ISM but also has
several point sources including M82 X-1. The total energy spectrum is very complex, but 
the line profiles contains plenty information about the physical status of hot and cool ISM.
 \citet{Strickland2007} pointed out the Fe I (6.4 keV) line, 
and a  velocity dispersion of a few 100 km$^{-1}$ in the Fe XXV (6.7 keV) line.  We simulated the energy 
spectrum based on \citet{Konami2011}. If the only hot (Fe XXV) ISM has a velocity dispersion 
of  200 or 500 km s$^{-1}$, it will be measured to 40\% accuracy.  If the cool ISM is static, while  the hot ISM is 
expanding, we have an indication that physical processes associated with the starburst activity transfer energy directly to a hot outflow. 
%it will indicate the physical process of energy transfer from starburst activity directly to the hot outflow. 
If, on the other hand, the cool and hot ISM interact directly, charge exchange emission can be observed, and 
%the line ratio between the K$_{\alpha}$ or Ly$_{\alpha}$ and others (so-called "G-ratio") will shift from 
%that of  thermal emission from  the ionization  equilibrium plasma. 
the intensity ratio between the forbidden plus intercombination lines to the resonance line in the He-like triplets (the `G-ratio') will differ from the range observed in collisional equilibrium plasmas.
Using the He-like Mg triplet, the G-ratio can be determined within a few \% accuracy (or limited by systematic errors)
by a 200 ksec SXS observation. 

\begin{figure}[htbp]
\begin{center}
\includegraphics[width=0.45\textwidth, bb=0 0 595 542]{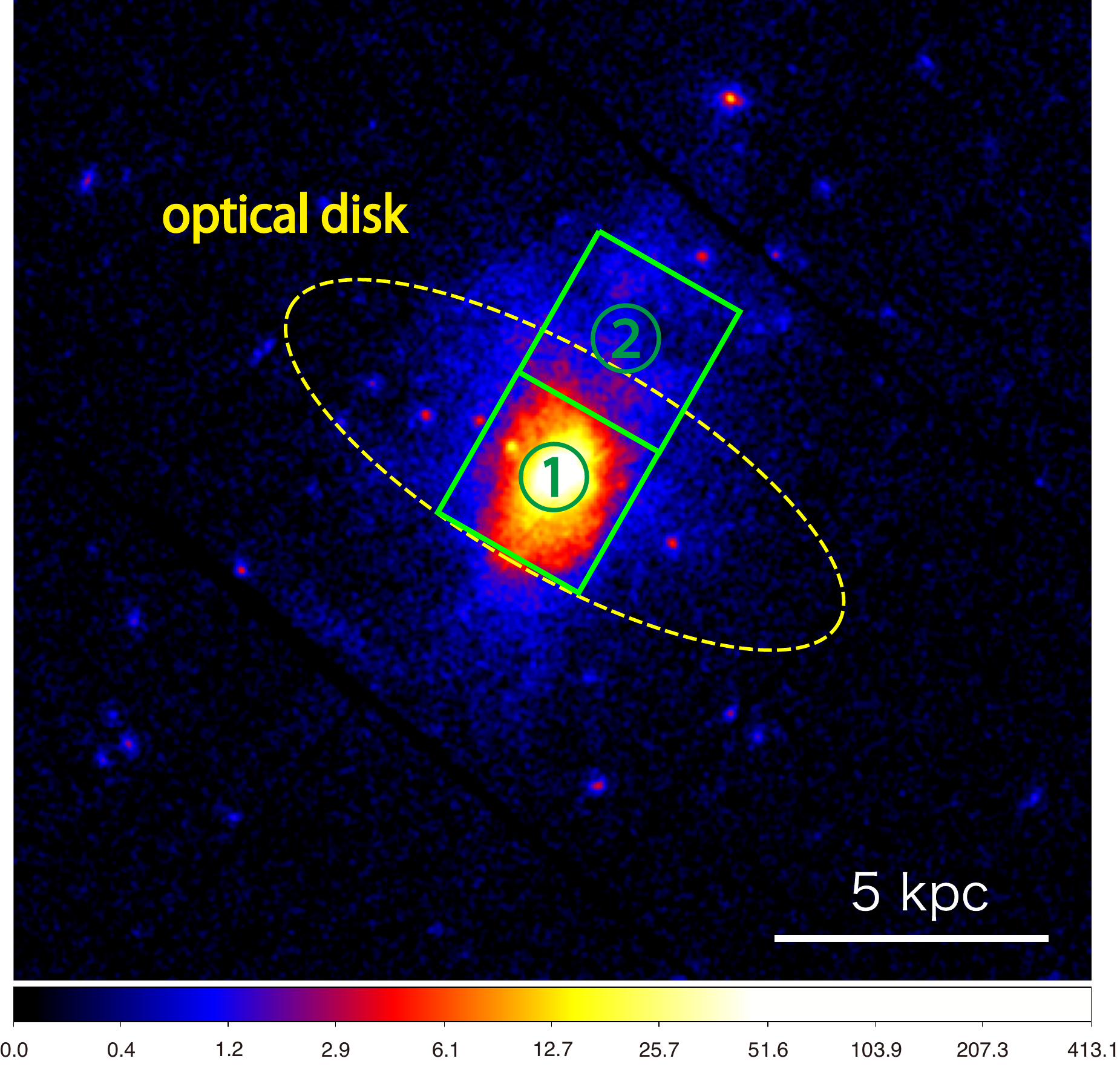}
\caption{The {\it XMM-Newton} image of M82. The two green squares correspond to the SXS field of view at 0 and 3 arcmin offset from the center. The yellow ellipse shows the optical disk.}
\label{fig:M82-image}
\end{center}
\end{figure}

In order to study the chemical composition of the hot ISM in starburst galaxies, we select the M82 halo, because 
it is brighter that that of NGC 253, and less complex than in the core region. We select a region just above the core
as shown in the figure \ref{fig:M82-image}, and call it the "Wind"  region. 
The stray light from the core shall be subtracted by the observed data. In a 200 ksec observation, the 
metal abundances can be determined by Fe L-line complex and the He-like Mg triplet to a 20\% accuracy.

\begin{figure}[htbp]
\begin{center}
\includegraphics[width=0.96\textwidth]{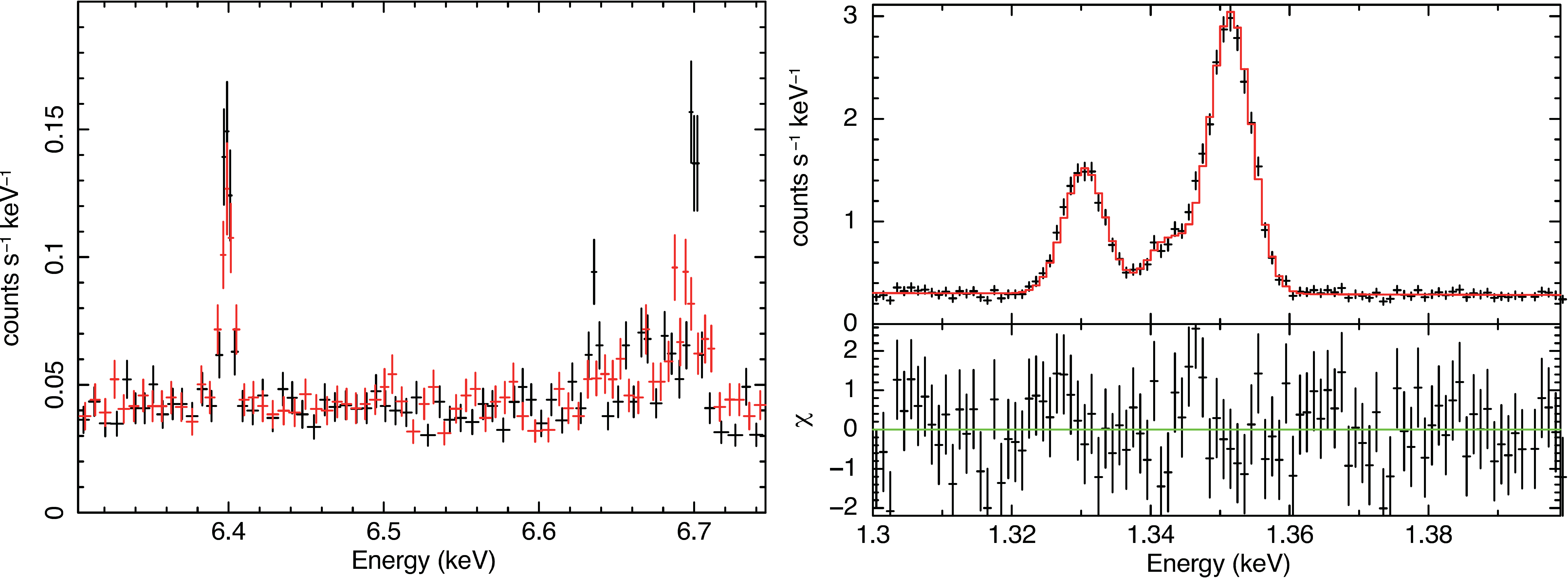}
\caption{(left) A simulated SXS spectrum assuming an energy resolution of 7 eV of the M82 center region with 200 ksec exposure. The He-like Mg triplet is shown on the right. The Fe K band spectrum (left) shows fluorescent Fe K at 6.4 keV and the He-like Fe XXV triplet at 6.7 keV. Black and red data show the He-like triplet lines with 0 km/sec and 500 km/sec velocity dispersions, respectively. }
\label{fig:M82-disk}
\end{center}
\end{figure}

\subsection{Beyond Feasibility}

Alternative methods include SXS spectroscopy of the hard X-ray line emission (in particular Fe XXV He$\alpha$; Strickland \& Heckman 2007) or absorption-line spectroscopy of background AGN or Ultra-Luminous X-ray sources (ULXs).    
A similar analysis on the circumgalactic gas of suitable starburst galaxies, using absorption spectroscopy on 
bright ULX's and emission line spectroscopy from the halo, will be discussed below.
We consider two cases, one using the ULXs in the outer galaxies, and the other is a QSO with an intervening absorption system.
One bright ULX, M82 X-1, is 2 orders of magnitude fainter than Mrk 421. As the column density of M82 
will be higher, emission from hot gas is difficult to eliminate from the observed spectrum of M82 X-1. Thus the Oxygen absorption lines 
will be hard to detect, and Ar or Si absorption lines may be better candidates. 

The QSO A0235 +094 ($z=0.94$) has an intervening cloud at $z=0.524$, observed by several satellites, with a column density of 2.4$\times10^{21}$cm$^{-2}$, as determined by {\it XMM-Newton} \citep{Raiteri2005}, which makes it an good candidate for spectroscopy with SXS, with a possible abundance determination for the intervening absorber.
Absorption features of redshifted Oxygen with a column density of 
$\sim$ 10$^{15}$  cm$^{-2}$ were detected in Mrk 421 at outbursts 
with {\it Chandra} LETG \citep{Nicastro2005} in 200 ksec exposure. The 0.5--2 keV flux of Mrk 421 in outburst 
reaches $10^{-9}$ erg cm$^{-2}$ sec$^{-1}$ at the peak. 
The SXS has a larger effective area of 50 cm$^{2}$ than {\it Chandra} LETG/ACIS ($\sim$ 5 cm$^{-2}$) in the Oxygen K band,
while the resolving power of $E/\Delta E$ of 120 (or $\Delta E$ = 5 eV) is smaller than that of LETG (20$\times\lambda \approx 400$). 
Thus the required exposure time will be, roughly, 1 Msec for a firm detection, using the approximate scaling with area and resolution of the sensitivity to weak absorption lines on a strong continuum.

%Background QSO: AO 0235+164; ULX's;
%Combine ULX absorption with halo emission: do same as for our own galaxy.

%{\bf TARGET: starburst with (1) sufficiently bright ULX for absorption spectroscopy (count rate
%scales very roughly as 0.1 (L/1E40) * (d/10 Mpc)$^{-2}$ counts/sec, (2) large (fills SXS FOV) bright halo (surface brightness several times higher than MWG)).}

%We would propose to retain the {\it goal} of measuring the outflow velocities directly in NGC 253 and M82, working closely with the calorimeter team to design an observation for which the gain uncertainty could be minimized. We may be able to contribute to knowledge of the gain calibration via our measurements.  
%The spectral range of interest is reasonably well
%sampled by the in-lab calibration lines (0.677, 1.04, 1.25, and 1.49 keV) may allow better energy-scale certainty there than available overall.  We would use
%the MXS Al calibration source on orbit to provide the lowest-energy calibration line and will evaluate the centroid of as many lines as possible from the source to constrain the uncertainties.  We will see to answer the question: "
%Does the shift scale with absolute energy (as would a real velocity) or with 
%dE from a calibration line (as would a calibration uncertainty)?"

%%%%%%%%%%%%%%%%%%%%%%%%%%%%%%%%%%%%%%%%%%%%%%%%%%%
\section{Emission Line Spectroscopy of Early-type galaxies}

%%%%%%%%%%%%%%%%%%%%%%%%%%%%%%%%%%%%%%%%%%%%%%%%%%%%%%%%%%%%
\subsection{Background and Previous Studies}
%%%%%%%%%%%%%%%%%%%%%%%%%%%%%%%%%%%%%%%%%%%%%%%%%%%%%%%%%%%

Early-type  galaxies have a hot, X-ray emitting 
ISM whose origin is considered to be the accumulation of stellar mass loss.
The timescale for the accumulation of mass-loss products is smaller than 1 Gyr
and the luminosities of the hot ISM in most early-type galaxies
are consistent with the energy being input from such stellar mass loss \citep{Matsushita2001}.
Recent supernovae (SN Ia) may inject additional metals, especially Si and Fe, and 
may also impart some energy to the ISM.
Some early-type galaxies have ISM luminosities
much higher than the rate of energy input from stellar mass loss.
These galaxies are located within cool cores of group--scale 
potential structures that have a  
higher gravitational mass in the outer regions
than the X-ray faint galaxies \citep{Matsushita2001, Nagino2009}.
As in cool-cores in massive clusters of galaxies, 
there should be some form of energy injection.
In these galaxies,
\cite{Allen2006} found
 a tight correlation  between the Bondi accretion rates 
 and the power emerging from these systems in relativistic jets. 
As a result, they are also suitable to study feedback from 
active galactic nuclei.

%\subsubsection{Measurements of turbulence using the effect of
% resonant line scattering}

The heating of the Intracluster Medium (ICM) may be driven by turbulent motions of the gas.
 Such turbulent motions in cool cores can be constrained by 
observing the effect of resonant scattering. Without turbulent
motions, the hot Interstellar Medium (ISM) in the center of  luminous 
early-type galaxies can be optically thick within some of
the strong Fe-L lines.
 \cite{Xu2002} found that 
the flux of the Fe XVII line at 0.83 keV of the central region
of an X-ray luminous elliptical galaxy, NGC 4636,  observed with {\it XMM-Newton} RGS
is smaller than what is expected from the APEC plasma code \citep{Smith2001}.
This discrepancy 
 indicates that resonant  scattering suppresses the line flux emerging from the central region
and the upper limit for the turbulent velocity in this galaxy
is lower than 200 km/s, assuming isotropic turbulence 
\citep{Xu2002,Werner2009}.
\cite{Werner2009} and \cite{dePlaa2012} also estimated the
effect of the scattering for several elliptical galaxies,
and found that the measured line ratios are different among galaxies.
Although there is a systematic uncertainty in the theoretical line ratio,
the difference in the measured values indicates a difference in the 
strengths of the turbulence.

%\subsubsection{Metal abundances in the hot ISM}

\begin{figure}[t]
 \includegraphics[width=1.0\textwidth]{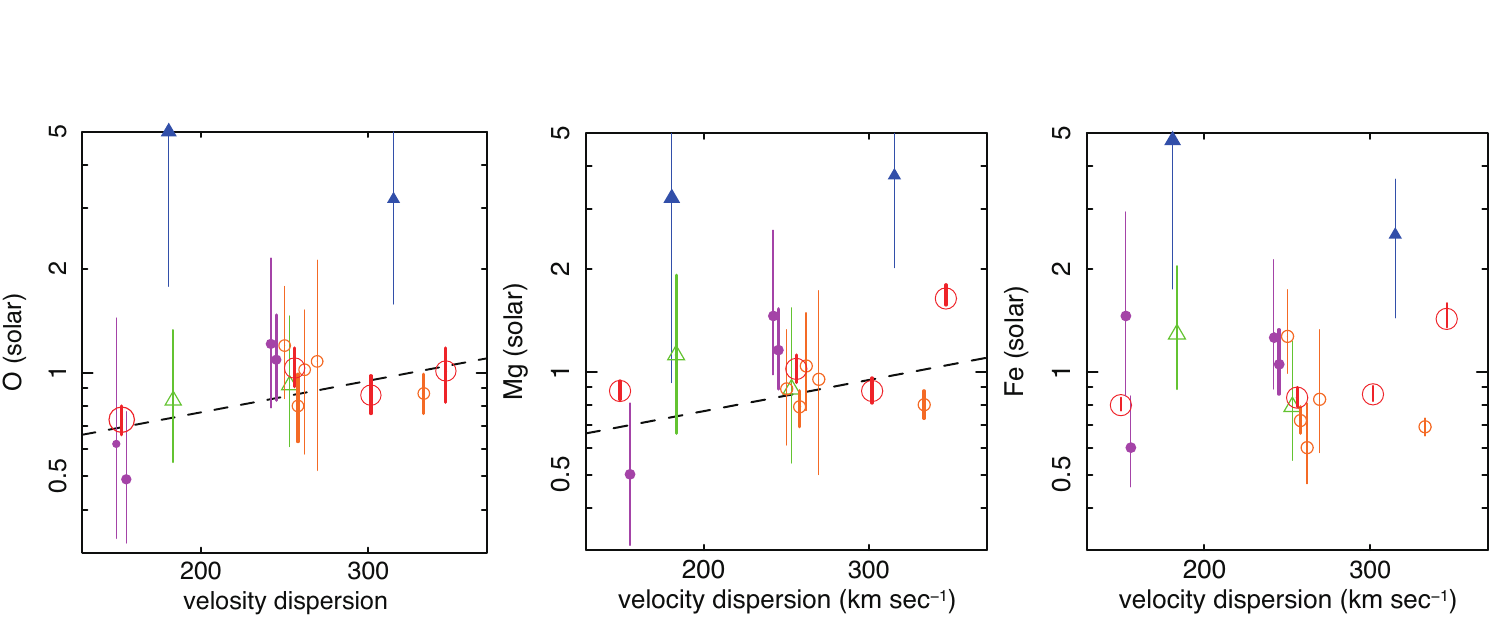}
\caption{The abundances of O, Mg, and Fe in the hot ISM in early-type
galaxies observed with {\it Suzaku} \citep{Konami2014},
plotted against the central stellar velocity dispersion, $\sigma$.
The dashed lines show the metallicity-$\sigma$ relation within 
about the half light radius aperture derived from optical spectroscopy
by \cite{Kuntschner2010}.
}
\label{img:suzakuabund}
\end{figure}

Measurements of 
stellar metallicity in galaxies  provide important information in 
understanding the history of star formation and the evolution of
galaxies.
Optical observations indicate that the abundance ratios of $\alpha$-elements
and Fe  of stars
are super-solar in the cores of bright early-type galaxies and that they increase
with the galactic mass \citep[e.g.][]{Kuntschner2010}. This overabundance of $\alpha$-elements relative to Fe is a key indicator
that galaxy formation occurred before a substantial number of SNe Ia
could explode and contribute to lowering these ratios.
However, absorption-line indices that account for abundance ratios also
depend on the age distribution of stars. 
In addition, optical spectroscopy is limited to about the central half light 
radius  of galaxies.

Using X-ray observations, we can both directly determine the metal abundances
of the ISM and constrain the stellar metallicity of the entire
galaxy. In particular, the O, Ne, and Mg abundances should reflect the stellar
metallicity, since these elements are predominantly synthesized in 
core-collapse SNe (SNecc).
The atomic data for lines at X-ray wavelengths are simpler than
for those in optical spectra, and the structure of the hot ISM is also
much simpler than stellar population data. 
Although some early-type galaxies show dust emission, the dust sputtering
time scale in the ISM is much shorter than the timescale for the ISM.
We can therefore estimate
the temperature and metallicity of the hot ISM through X-ray spectra
with small systematic uncertainties, after calibrating the systematic uncertainties in the Fe-L atomic data.
In addition, because Ne is not injected into the ISM in the form of dust, 
to reduce the systematic  uncertainties caused by the existence of dust we need 
measurements of Ne in the ISM.

\begin{figure}[t]
\begin{center}
 \includegraphics[width=0.4\textwidth]{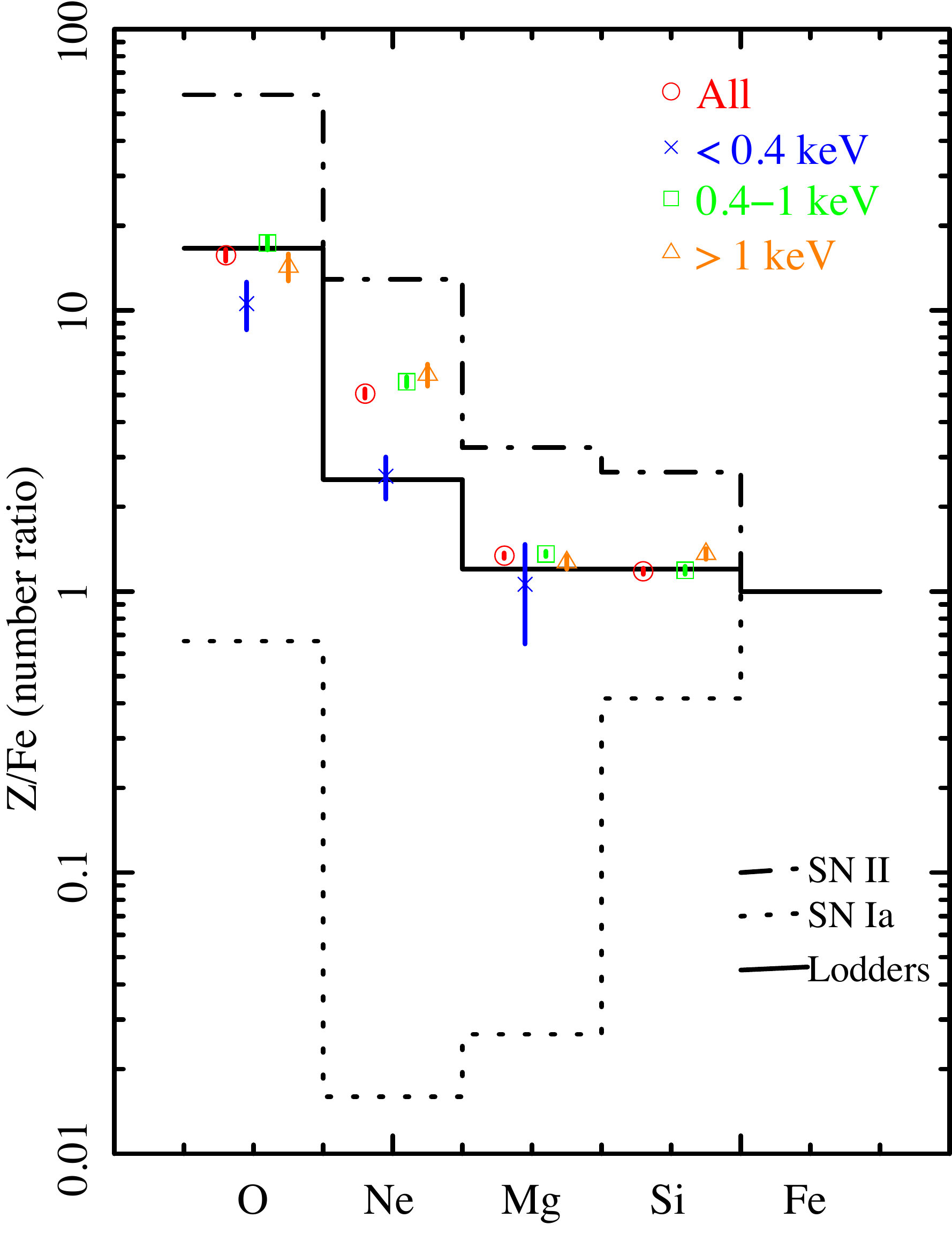}
 \includegraphics[width=0.45\textwidth]{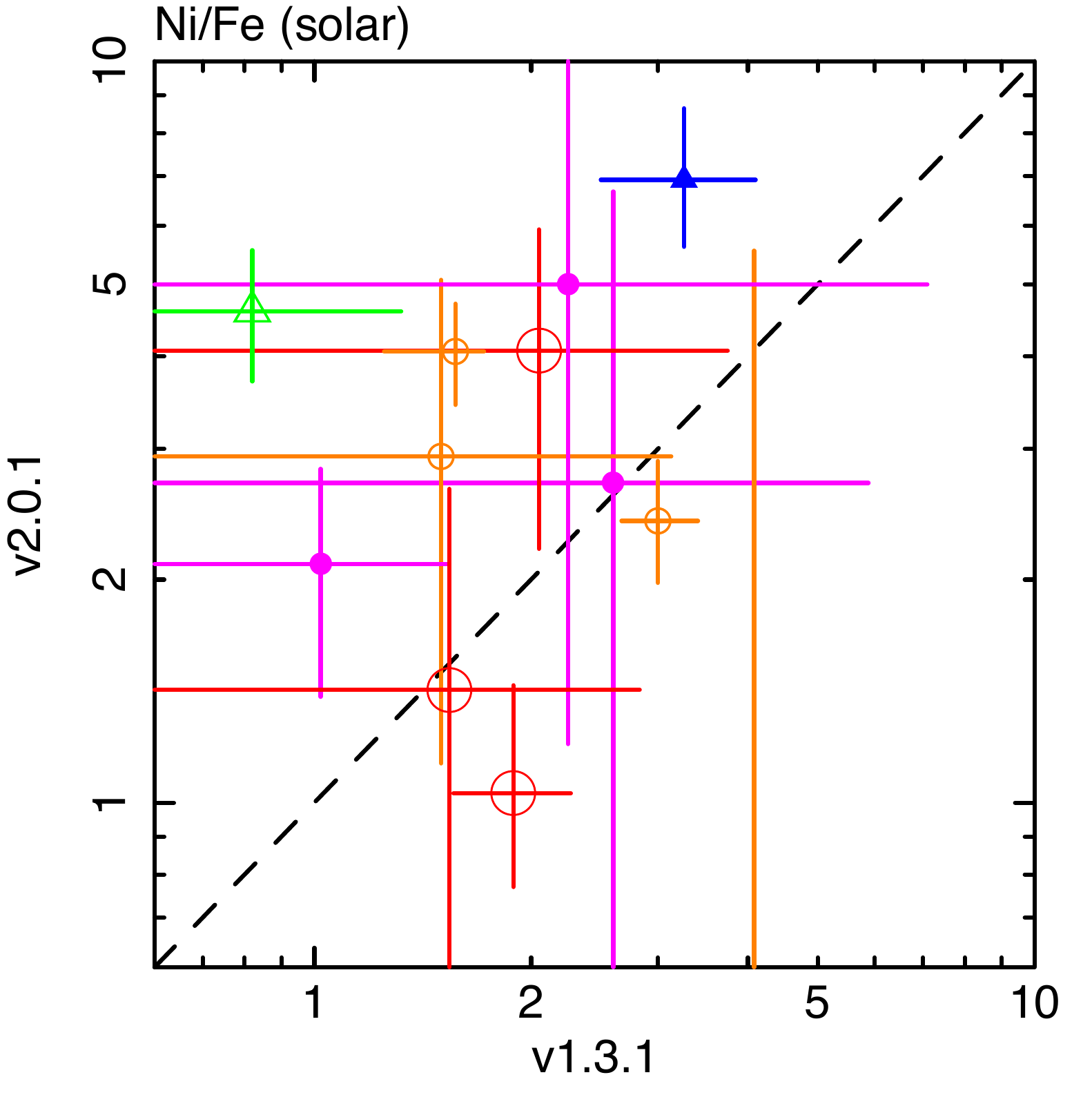}
\caption{(left panel)
The weighted averages of the abundance ratios of O, Ne, Mg, and Si to Fe
for 17 early-type galaxies observed with {\it Suzaku} \citep{Konami2014}, derived with APEC v2.0.1.
Red, blue, green, and orange data show for all the sample galaxies,
 temperature groups of $<$0.4 keV, 0.4-1 keV, and $>$1 keV,
 respectively. Solid, dot-dashed, and dotted lines represent the number
 ratios of metals to Fe for solar abundance \citep{lodd2003}, for SNcc
 products \citep{Nomoto2006}, and for SN Ia ones \citep{Iwamoto1999}, respectively.
(right panel) Comparison of the Ni/Fe ratios in units of the solar ratio
in the ISM of early-type galaxies observed with {\it Suzaku},
derived with APEC v1.3.1 and v2.0.1. 
The dashed line indicates equal value between two models.
}
\label{img:suzakuabundneni}
\end{center}
\end{figure}

The metal abundances in the ISM have been studied with {\it ASCA}, {\it Chandra},
{\it XMM-Newton} and {\it Suzaku}.
Early measurements of the ISM with {\it ASCA} showed that metallicity is 
less than half of solar abundances, which is significantly smaller
than the stellar metallicity, and the expected contribution from SN Ia
  \citep[e.g.][]{Awaki1994}.
The uncertainties in the temperature structure and/or in the Fe-L atomic data
sometimes yield very large systematic uncertainties in the derived abundances
in the ISM, especially from CCD spectra \citep[e.g.][]{Matsushita2000}.
Using plasma code with revised atomic data, as shown in Figure \ref{img:suzakuabund},
  \citet{Konami2014} found that
the derived O and Mg abundances in the ISM with {\it Suzaku} mostly agree with 
a metallicity-$\sigma$ relation derived from optical spectroscopy
within about  the effective radius, or about the
 half light radius.
 In contrast, Figure \ref{img:suzakuabundneni} shows that
the derived Ne/Fe ratios are about a factor 
of two higher than  the O/Fe, Mg/Fe, and Si/Fe ratios in units of the solar ratio.
The derived Ni/Fe ratios become 1--8 in units of the solar ratio, 
with a very large scatter, and the difference in the Ni/Fe ratios
between the two versions of APEC codes  sometimes reaches a factor of 5 (Figure \ref{img:suzakuabundneni}).
These large Ne/Fe and Ni/Fe ratios cannot be explained by any nucleosynthesis
models for SNcc and SN Ia.
Since Ni-L  and Ne-K lines are hidden in the Fe-L lines,
the abundances of these elements are derived  from residuals between 
the data and the Fe-L model, which strongly depends on the plasma codes.
These anomalies in the Ne and Ni abundances therefore may be caused by remaining 
systematic uncertainties in the plasma codes.

The Fe abundance in the ISM is a sum of stellar metallicity and the contribution 
from SN Ia.
The standard SN Ia rates derived from optical observations predict the
Fe abundance in the ISM to be as high as several times the solar value
\citep[see][in detail]{Konami2014}.
However,
the derived Fe abundances in the ISM with {\it Suzaku} are about 0.8 solar (Figure \ref{img:suzakuabund}).
Subtracting the contribution from SNecc, the Fe abundance from the SN Ia
becomes $\sim$ 0.5 solar.
Therefore, if all the ejecta of SNe Ia have been completely mixed into the ISM, the present SN Ia rate 
needed to account for the observed Fe abundance in the ISM is significantly smaller than what
is  expected.
The ICM in clusters of galaxies contains a large amount of Fe and 
 the metal abundances in the ISM in early-type galaxies provide important information about the present metal supply into the ICM through SN Ia and stellar mass loss.
 However,
accumulating the present supply of Fe derived from the Fe abundance in the ISM,
the resultant Fe mass is lower by two orders of magnitude than
that in the ICM.
One possibility is that the lifetimes of most SNe Ia are much shorter than the
Hubble time, and the SN Ia rate in cluster galaxies was much higher in the past.
Another possibility is that
 if some part of SN Ia ejecta can escape the ISM before being
fully mixed into the ISM, the Fe abundance can be lower. 
Furthermore, there may remain systematic uncertainties in the abundance 
measurements caused by uncertainties in the Fe-L atomic data.

\subsection{Prospects \& Strategy}

The X-ray spectra from the hot ISM in early-type galaxies are dominated
by very complicated Fe-L emission.
Severe discrepancies still remain between plasma codes which
calculate the X-ray emission spectra of an optically thin plasma.
The energy resolution of the {\it ASTRO-H} SXS in the Fe-L energy band is significantly better
than the RGS onboard {\it XMM-Newton} for extended sources whose spatial scale
is larger than  $1'$.
For example, at 1 keV, the spectral resolving power for a extended source with $1'$ source extent  is about 100 using the RGS detectors.
Calibration of the plasma codes using the {\it ASTRO-H} SXS data will therefore
be critically important to study the effects of the resonant scattering
and abundance measurements in the ISM.
The ISM in these galaxies is expected to be close to ionization equilibrium,
and the temperature structure is relatively simple, the density is low and
the ISM is mostly optically thin. 
As a result, 
the ISM in early-type galaxies is much simpler than the X-ray emitting regions
of stars, supernova remnants, hot gaseous halos in spiral galaxies,
and cool cores in clusters of galaxies.
They are thus the best target to calibrate the Fe-L emission, which is very important
to the study of other SXS spectra dominated by Fe-L lines.

With the {\it ASTRO-H} SXS, 
we can study the spatial variation of the line ratios of
optically thick and optically thin lines and study the effects of resonant 
scattering, though direct measurements of turbulent velocities are challenging.
For example, for NGC~4636, the expected optical depth is about 9 for
the 0.83 keV Fe XVII line and about unity for Fe XVIII line at 0.87 keV
and Ly$\alpha$ O line \citep{Werner2009, Churazov2010}.
Thus, the flux ratios of the 0.83 keV line and optically thin lines
with the same ionization (for example, Fe XVII lines at $\sim$0.73 keV)
are usually used to study the effect of the scattering.
Without turbulent motions, the  0.83 keV line flux from NGC~4636 is expected to be 
modified out to a few arcminutes, which will be resolved with {\it ASTRO-H}.
To calibrate the atomic data,
 offset observations,  where both the
temperature structure is simpler than in the core regions and the 0.83 keV line 
becomes optically thin, are important.

Observations of the hot ISM in early-type galaxies 
enable us to study the present metal supply from early-type galaxies.
Thus, observations of these galaxies with the {\it ASTRO-H} SXS
are complementary to those of clusters of galaxies, since
the metals in the ICM in the cool core of clusters are a
mixture of those present in the ICM and those supplied later from their cD galaxies.
Outside the cool-cores, the metals accumulate
over much longer time scales.
With the SXS,  the Ni and Ne lines will become free from the complex Fe-L lines 
and we will be able to derive abundances of these elements with much smaller
systematic uncertainties.
Furthermore, with superior energy resolution and a modest effective area,
we will able to detect very faint line emission such as Al-K lines.
Nucleosynthesis models indicate that the abundances of
odd-Z elements like Al  show a strong dependence of 
stellar metallicity \citep{Nomoto2006, Kobayashi2006}.
These elements are enhanced by the surplus of neutrons in $^{22}$Ne,
which is synthesized by CNO cycle during He-burning.
In this way, the abundance  of Al in the ICM
can be used as an indicator of the stellar metallicity.

Although most of the metal is provided by SNe or massive stars and white dwarfs, C and N are also 
ejected from low-mass and 
intermediate-mass stars during the asymptotic
giant branch phase.
The formation history of intermediate mass stars is poorly known, compared to 
massive stars which emit strong optical emission lines and low mass stars
whose life time is very long.
The abundance measurements of C and N in the ISM will give us some
information of history of the intermediate mass stars.

If present SNe Ia ejecta are not fully mixed into the ISM,
they may remain very hot and, due to their buoyancy, may be escaping
from galaxies \citep{Matsushita2000, Tang2010}.
Fe-K line emission from these ejecta may then be observed with the SXS.
Even if the Fe-K line is not detected, we can place tight constraints
on the metal supply from present SNe Ia in these galaxies, and hence
constrain the Fe enrichment history in clusters of galaxies.

\subsection{Targets \& Feasibility}

\begin{figure}[tb]
\begin{center}
 \includegraphics[width=0.45\textwidth]{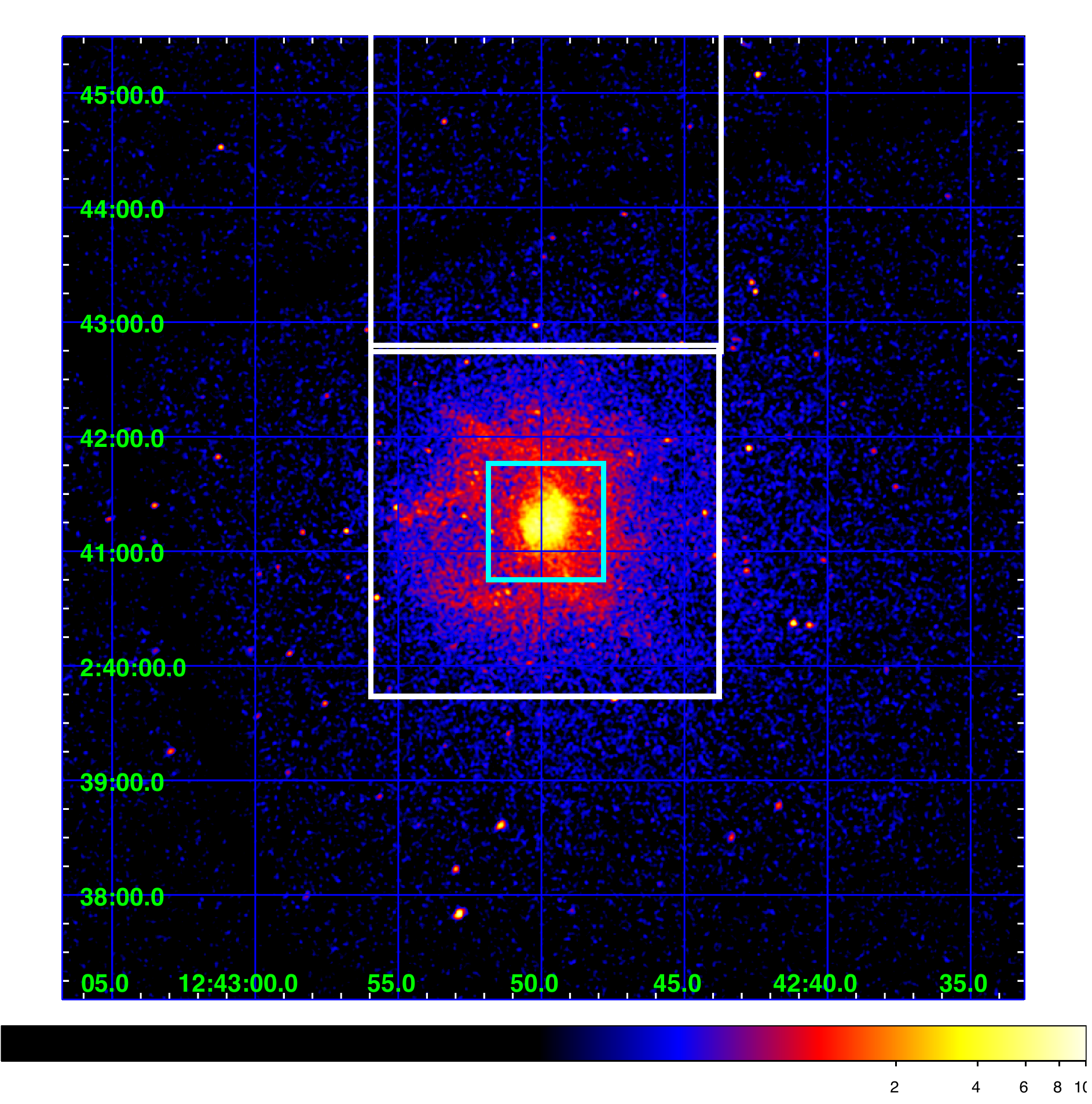}
 \includegraphics[width=0.45\textwidth]{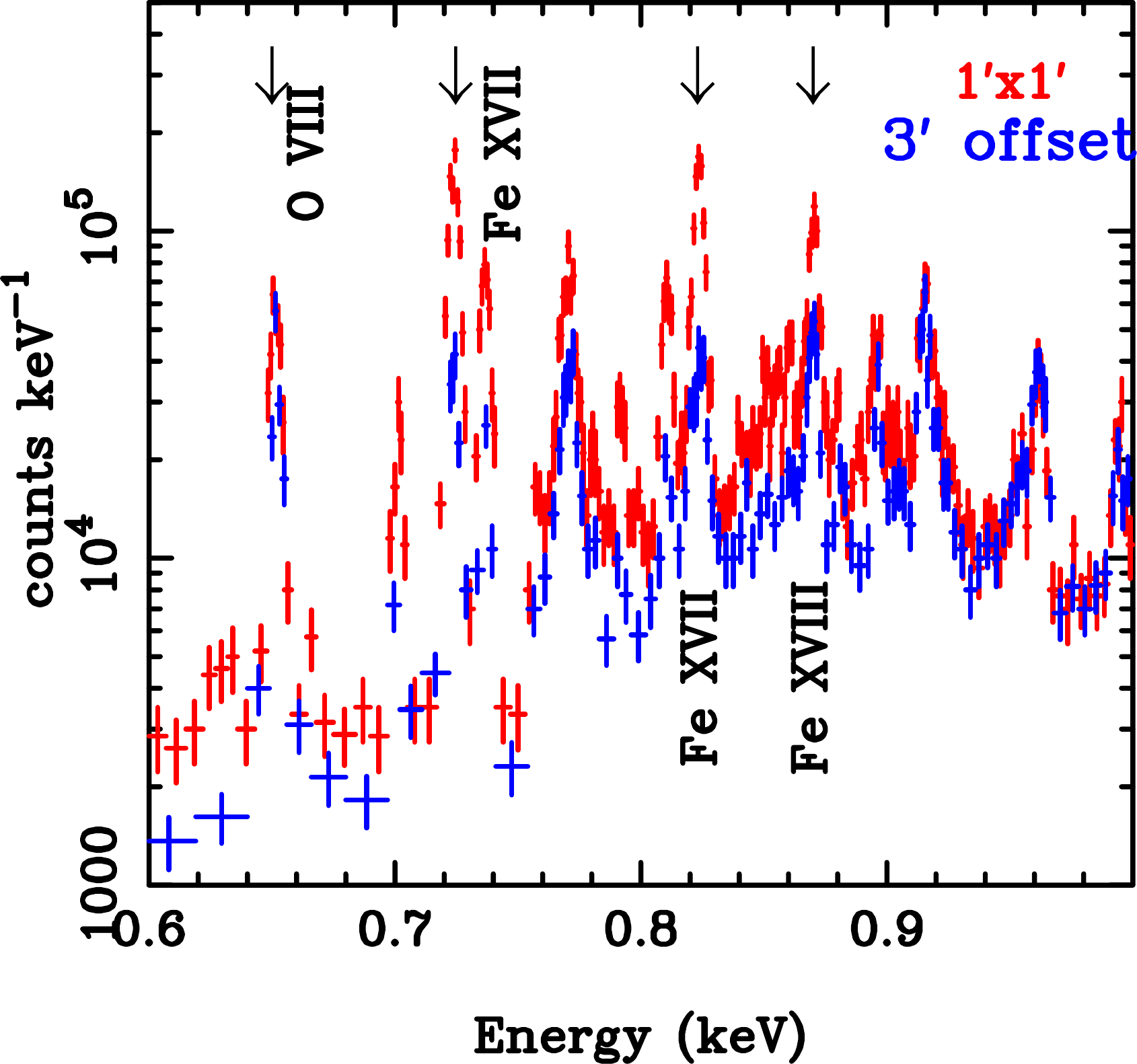}
\caption{(left panel) The {\it Chandra} image of NGC 4636. The two white squares
correspond to the SXS field of view at $0'$ and $3'$ offset from the
center. The cyan square shows the central $1'\times1'$ pixels.
(right panel)  Simulated SXS spectra of NGC~4636 at the Fe-L line energy
 range for the central $1'\times 1'$ (red) and $3'$ offset
 pointing (blue).
The exposure time for each pointing is 100 ks. 
The expected optical depth for 0.83 keV Fe XVII,  0.87 keV Fe XVIII, 
0.65 keV OVIII, and 0.73 keV Fe XVII 
lines are about 9, 1.3, 1.2, and 0.5, respectively \citep{Werner2009}.}
\label{img:chandra}
\end{center}
\end{figure}

Located in the Virgo Southern Extension, NGC~4636 is one of the brightest giant
elliptical galaxies in X-ray wavelengths.
{\it ASCA} observations have shown very extended X-ray emission with a
radius at least 300 kpc surrounding this galaxy \citep{Matsushita1998}.
Using {\it Chandra} data \cite{Jones2002} found  arm-like structures 
as shown in Figure \ref{img:chandra}. They suggest
that these structures are
produced by shocks driven by symmetric off-center outbursts of the
central AGN. 
In spite of these  structures, the temperature structure seems relatively simple; the projected temperature profile
in the ISM with {\it Chandra} and {\it XMM-Newton} shows a relatively small variation \citep[e.g.][]{Werner2009}, and the RGS and {\it Suzaku} spectra are fitted with a single-temperature model.
As a result, this galaxy is one of the most suitable targets
to calibrate the Fe-L emission and to 
study the effects of the resonant scattering and abundance measurements.

We extracted MOS spectra of annular regions ($0.0'-0.5'$, $0.5'-1.5'$,
$1.5'-4.5'$, $4.5'-8.5'$) centered on NGC~4636 and fitted each spectrum
with a single-temperature vAPEC model.
We then used simx-1.3.1 to simulate SXS spectra with the 5 eV
response file, assuming 50 ks, 100 ks, and 200 ks exposures,
pointing at the center of the galaxy and at a position $3'$ offset toward north,
as shown in Figure \ref{img:chandra}.

The FOV of the $3'$ offset pointing is almost free from the effects of
resonant line scattering and  the arm-like shock structure,
and probably has a simple temperature structure.
The fraction of scattered photons due to the point spread function
(PSF) is about 15\% of detected photons within the FOV and is not
 significant.
The SXS spectrum for this region may therefore be used to evaluate
the validity of the atomic data.

\subsubsection{Resonant line scattering}

The right panel of Figure \ref{img:chandra} shows the 
simulated SXS spectra at the Fe-L energy range.
The expected optical depth of the 0.83 keV and 0.87 keV lines at the center of
NGC 4636 is about 9 and unity, respectively \citep{Churazov2010}.
Table \ref{counts} summarizes the expected number of photons
for a 100 ks exposure for the two Fe-L lines at 0.83 keV (Fe XVII) and
 0.87 keV (Fe XVIII).
 As shown in Figure \ref{img:chandra},
 we expect a similar number of photons from the optically thin Fe XVII lines
 at 0.73 keV.
The effect of line scattering for the 0.83 keV line is strongest 
within the central few kpc, which corresponds to about 0.5$'$.
Considering the effect of the PSF, about 60\% of the detected photons within 
the central $1'\times 1'$ region come from the $r<0.5'$ circular 
region on the sky.
Outside the central $1'\times 1'$ region, about 20\% and 70\% of photons
will come from $r<0.5'$ and 0.5$'<r<1.5'$ regions, respectively.
This demonstrates that we will able to constrain the amount of turbulence
by using information derived from the radial profile of the flux ratio of these lines 
to the other Fe-L lines. We will also be able to
evaluate the systematic uncertainty in the plasma codes.

\begin{table}[h]
\caption{Expected number of photons with an 100 ks exposure 
for the 0.83 keV and 0.87 keV Fe-L lines } 
\label{counts}
\begin{center}
\begin{tabular}[t]{ccccc}\hline
           & region  &  0.83 keV    & 0.87 keV\\\hline
\multicolumn{2}{c}{expected optical depth$^a$} & 8.8 &  1.3 \\\hline
$0'$ offset & central $1'\times 1'$ & 1000 & 700 \\
$0'$ offset & outside central $1'\times 1'$ & 2600 & 2100 \\
$3'$ offset & FOV  & 300 & 300\\\hline
\end{tabular}\\
\end{center}
$^a$: \citet{Churazov2010}

\end{table}

\subsubsection{Metal abundances in the ISM}

\begin{figure}[tb]
\centerline{
 \includegraphics[width=0.55\textwidth]{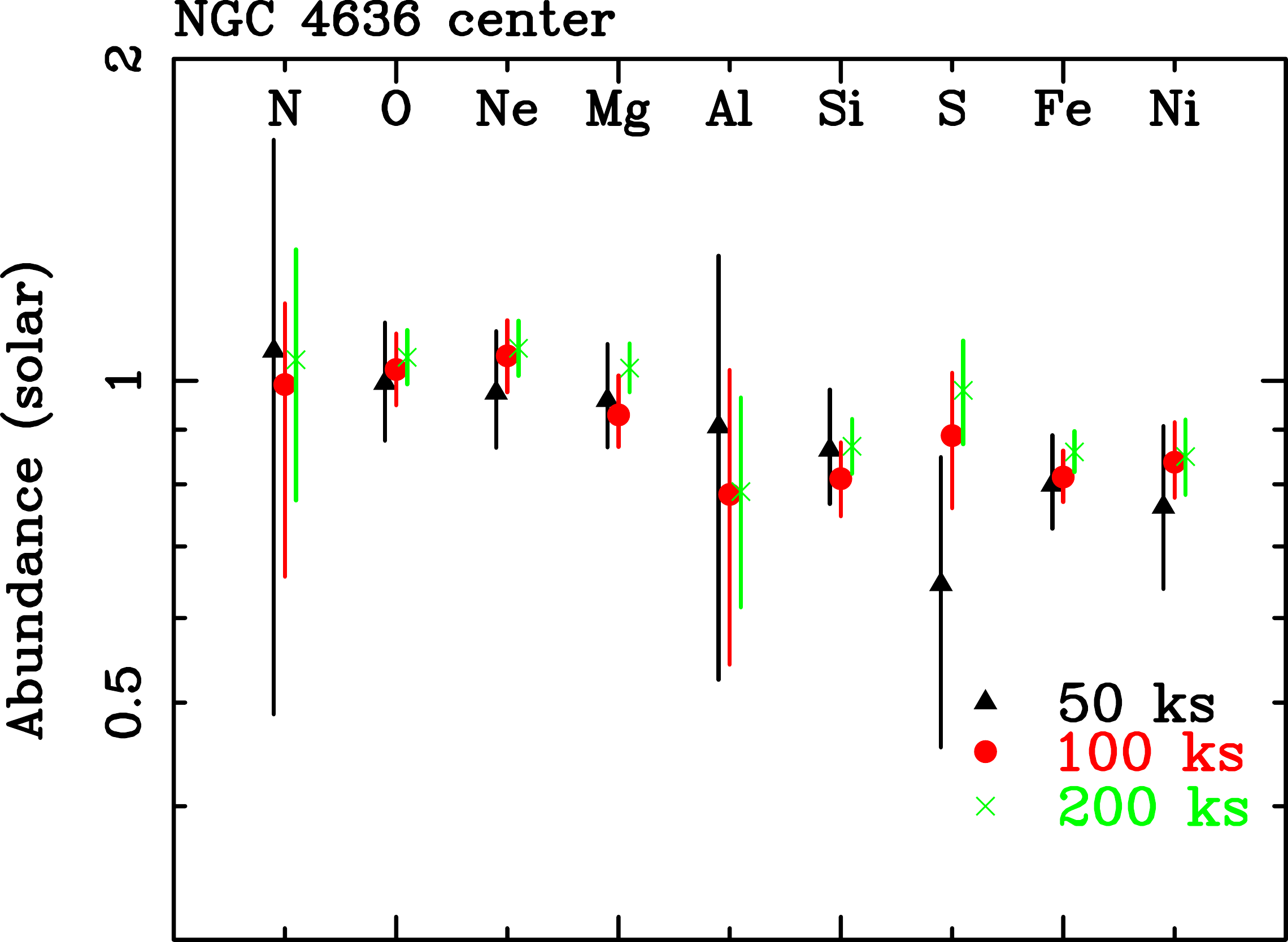}
}
\caption{ Abundances measured from the simulated SXS spectra of the 
central region ($3'\times3'$) of NGC~4636 assuming 50 ks, 100 ks, and
 200 ks exposures. Here, errors are quoted with 90\% confidence.}
\label{img:sxsabundance}
\end{figure}

\begin{figure}[tb]
\centerline{
 \includegraphics[width=0.45\textwidth]{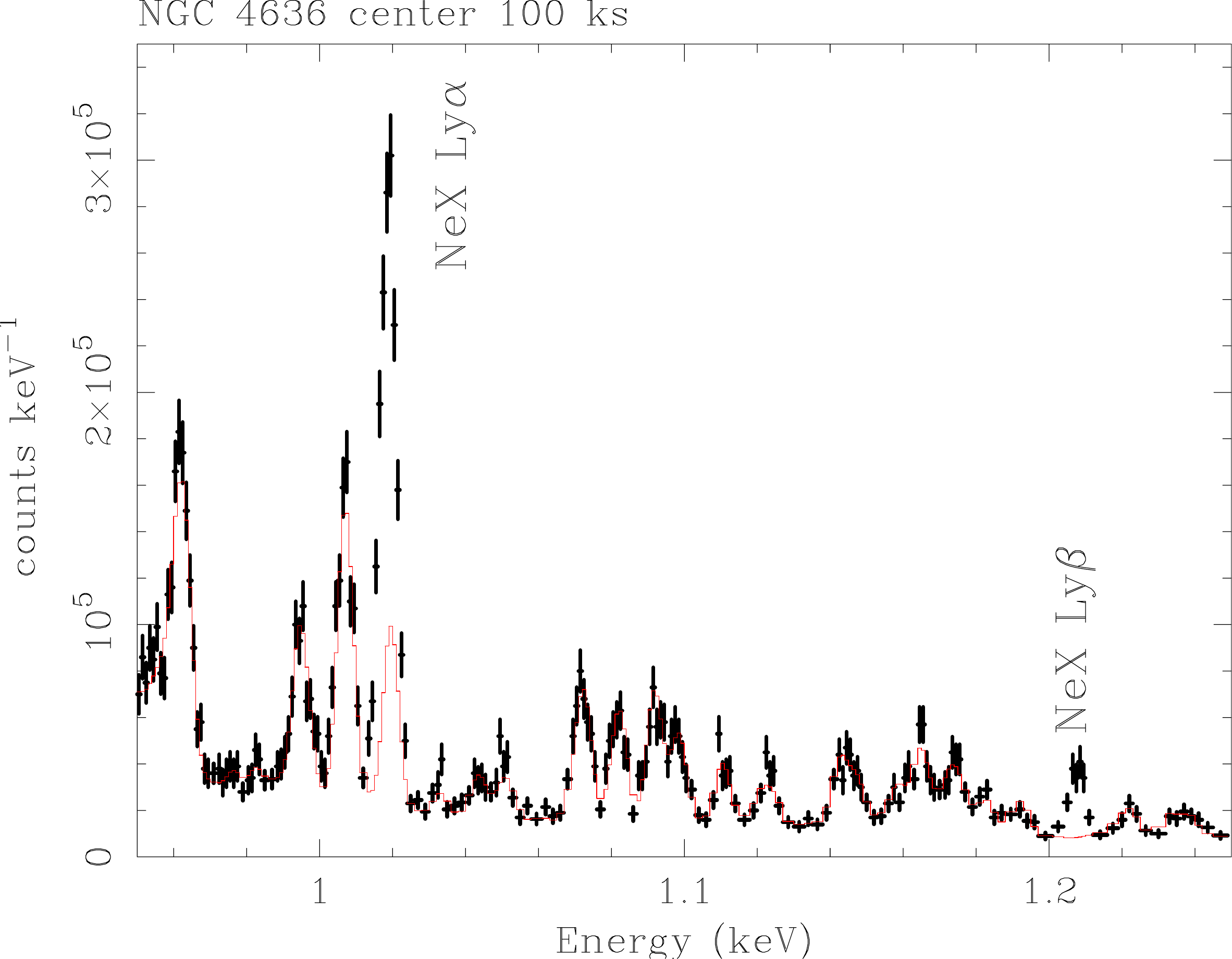}
 \includegraphics[width=0.45\textwidth]{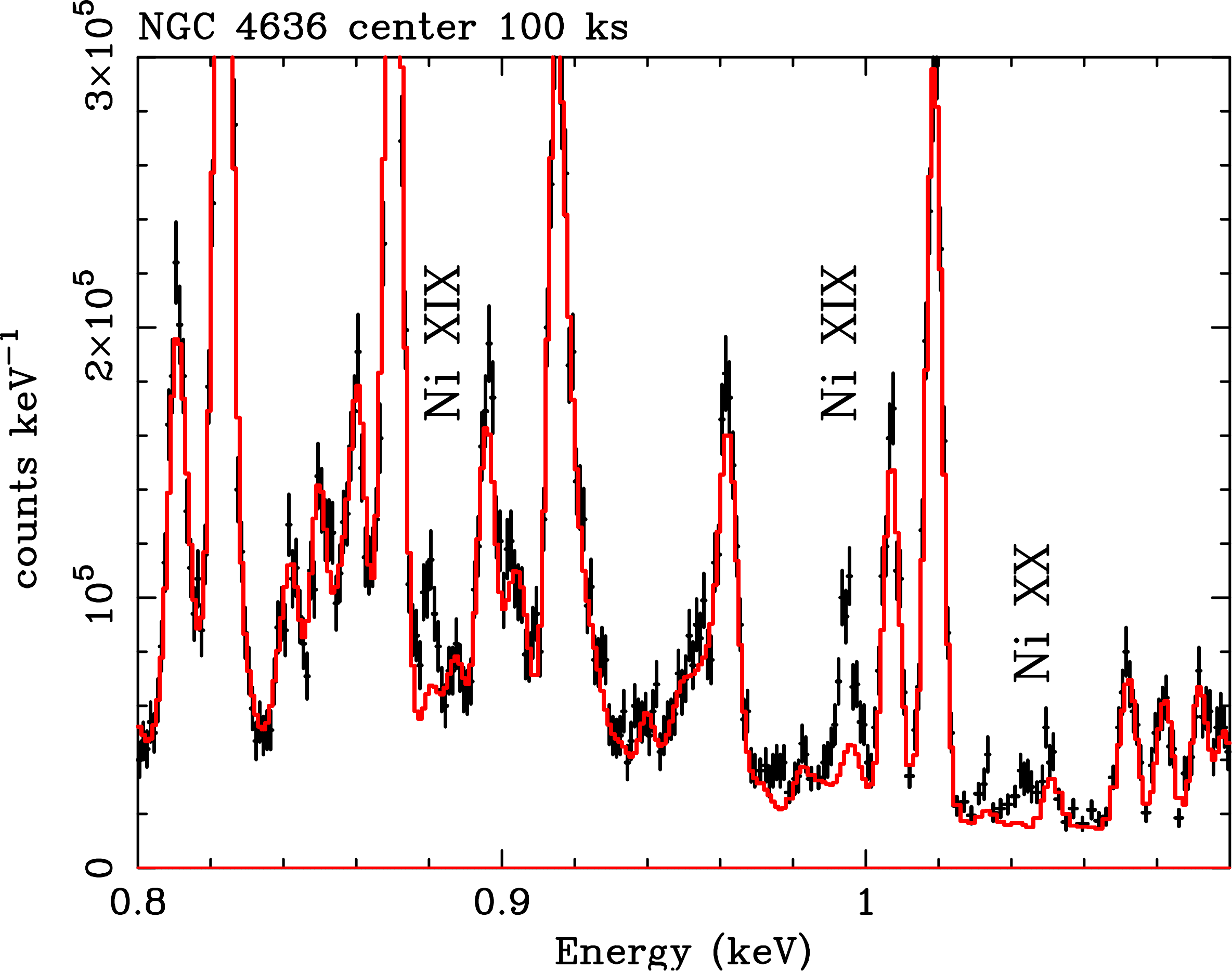}
}
\caption{The simulated SXS spectra of the 
central region ($3'\times3'$) of NGC~4636 assuming an exposure time of
 100 ks. The red lines in the left and right panel correspond to the
zero Ne and Ni abundance, respectively.}
\label{img:neni}
\end{figure}

Figure \ref{img:sxsabundance} shows the expected metal abundances
from the simulated SXS spectra of the central region of NGC~4636
assuming 50 ks, 100 ks, and 200 ks exposures.
From the central pointing,
abundances of O, Ne, Mg, Si, S, Fe and Ni in the ISM 
will be measured with a
10--20\% statistical accuracy with a 100 ks exposure.
In addition, the N abundance will be measured to an accuracy of 20--30\%.
Figure \ref{img:neni} shows the Ne and Ni lines in the Fe-L lines.
With the SXS, we can directly measure the strengths of the Ne-K and Ni-L
lines.
In particular, some lines such as the Ly$\beta$ line of Ne at $\sim$1.2 keV
and the Ni XIX line at 1 keV are mostly free from the Fe-L lines.

\begin{figure}[thb]
\begin{center}
 \includegraphics[width=0.55\textwidth]{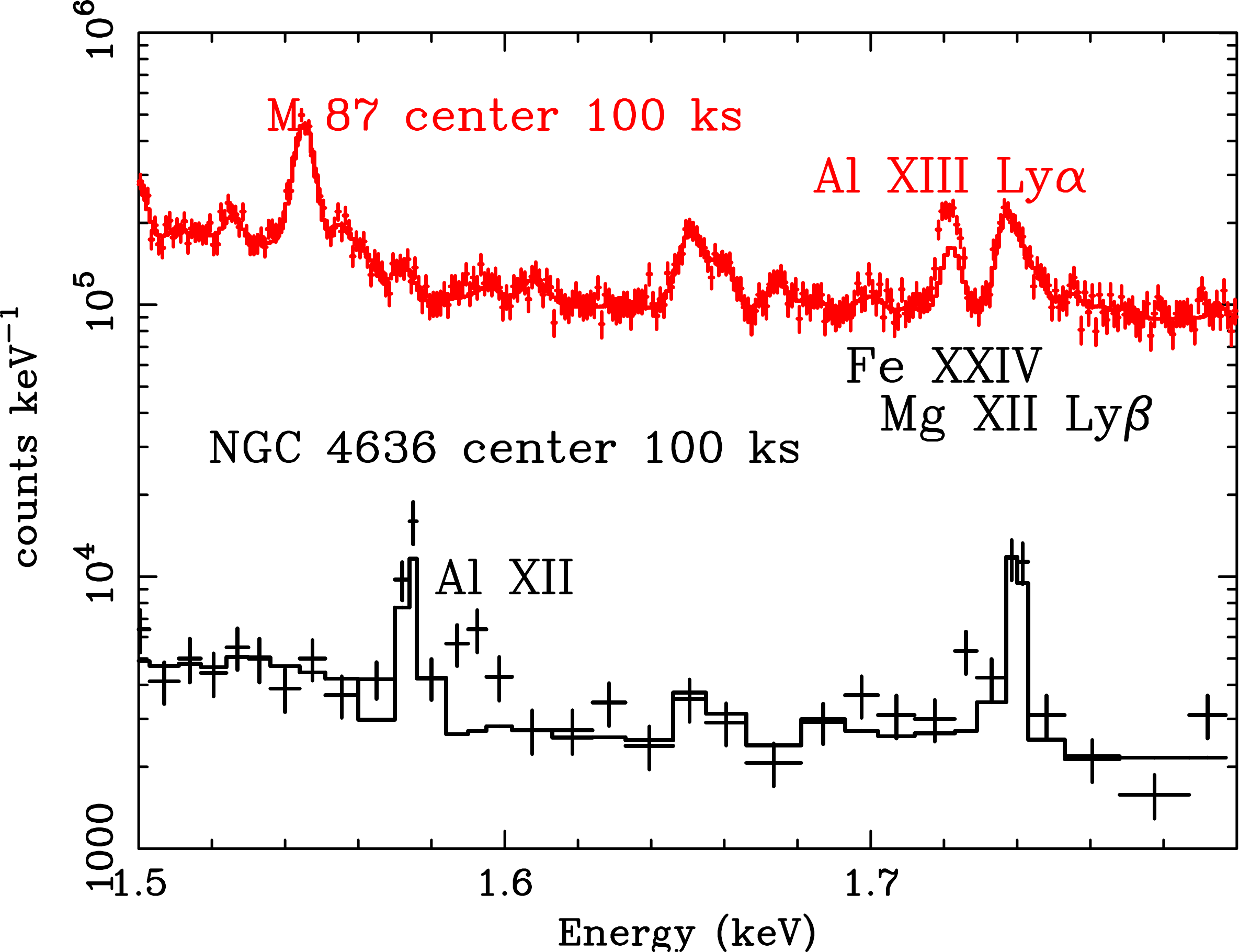}
\caption{The simulated SXS spectra at the  center of
NGC 4636 (black) and M 87 (red) with  100 ks exposures
around Al lines. Here, Al abundances are assumed to have 
the same values with the Mg abundances.
The solid lines correspond to zero  Al abundance.}
\label{nnaal}
\end{center}
\end{figure}

The Al abundance in the ISM of the central $3'\times3'$ region
will be measured with statistical errors of 0.4 solar, 
0.24 solar, and 0.17 solar, respectively, assuming 50 ks, 100 ks,
and 200 ks, exposures. 
Unfortunately, there are Fe  lines at almost the same energies as
 the Ly$\alpha$  line of Al (Figure \ref{nnaal}).
Here, we also show a simulated SXS spectrum  of the center of M 87
with a 100 ks exposure.
Although the statistics for NGC 4636 are much poorer
than that for M~87, the Al lines from NGC~4636 have less contamination
from the Fe-L line. Therefore, the systematic uncertainty caused by 
uncertainty in the atomic data should be much smaller in NGC 4636.

If one half of the Fe remains as hot (several keV) SNe Ia ejecta escaping from the ISM,
we expect several tens of  counts in the Fe-K line  with the SXS
from the central observation of NGC 4636 with a 100 ks exposure.

%%%%%%%%%%%%%%%%%%%%%%%%%%%%%%%%%%%%%%%%%%%%%%%%%%
%\section{References}
%%%%%%%%%%%%%%%%%%%%%%%%%%%%%%%%%%%%%%%%%%%%%%%%%%
%---------------------------------------------
% reference
%---------------------------------------------
\clearpage
\begin{multicols}{2}
\end{multicols}

\end{document}